\documentclass[10pt]{iopart}

\usepackage{graphicx}
\usepackage{dcolumn}
\usepackage{bm}
\usepackage{color}
\usepackage{cancel}

\begin{document} 

\title{Effect of the exoplanet magnetic field topology on its magnetospheric radio emission}

\author{J. Varela}
\ead{rodriguezjv@ornl.gov}
\address{Oak Ridge National Laboratory, Oak Ridge, Tennessee 37831-8071, USA}
\author{V. R\'eville}
\address{UCLA Earth, Planetary and Space Sciences, 595 Charles Young Drive East, Los Angeles CA 90095-1567}
\author{A. S. Brun}
\address{Laboratoire AIM, CEA/DRF – CNRS – Univ. Paris Diderot – IRFU/DAp, Paris-Saclay, 91191 Gif-sur-Yvette Cedex, France}
\author{P. Zarka}
\address{LESIA \& USN, Observatoire de Paris, CNRS, PSL/SU/UPMC/UPD/UO, Place J. Janssen, 92195 Meudon, France}
\author{F. Pantellini}
\address{LESIA, Observatoire de Paris, CNRS, UPMC, Universite Paris-Diderot, Place J. Janssen, 92195 Meudon, France}

\date{\today}

 
\begin{abstract}
     \textit{Context:} The magnetized wind from stars that impact exoplanets should lead to radio emissions. According to the scaling laws derived in the solar system, the radio emission should depend on the stellar wind, interplanetary magnetic field, and topology of the exoplanet magnetosphere. \\
     \textit{Aims:} The aim of this study is to calculate the dissipated power and subsequent radio emission from exoplanet magnetospheres with different topologies perturbed by the interplanetary magnetic field and stellar wind, to refine the predictions from scaling laws, and to prepare the interpretation of future radio detections. \\
     \textit{Methods:} We use the magnetohydrodynamic (MHD) code PLUTO in spherical coordinates to analyze the total radio emission level resulting from the dissipation of the kinetic and magnetic (Poynting flux) energies inside the exoplanet's magnetospheres. We apply a formalism  to infer the detailed contribution in the exoplanet radio emission on the exoplanet's day side and magnetotail. The model is based on Mercury-like conditions, although the study results are extrapolated to exoplanets with stronger magnetic fields, providing the lower bound of the radio emission.\\
      \textit{Results:} The predicted dissipated powers and resulting radio emissions depends critically on the exoplanet magnetosphere topology and interplanetary magnetic field (IMF) orientation. The radio emission on the exoplanet's night and day sides should thus contain information on the exoplanet magnetic field topology. In addition, if the topology of an exoplanet magnetosphere is known, the radio emission measurements can be used as a proxy of the instantaneous dynamic pressure of the stellar wind, IMF orientation, and intensity.
\end{abstract}


\maketitle

\ioptwocol


\section{Introduction}

The planets of the solar system and the exoplanets with intrinsic magnetic fields are emitters of cyclotron microwave amplification by stimulated emission of radiation (MASER) emission at radio wavelengths \cite{Kaiser,Zarka5}. This radio emission is generated by energetic electrons (keV) traveling along  magnetic field lines, particularly in the auroral regions \cite{Wu}, accelerated in the reconnection region between the interplanetary magnetic field (IMF) and the intrinsic magnetic field of the exoplanet. The magnetic energy is transferred as kinetic and internal energy to the electrons (a consequence of the local balance between the Poynting flux, enthalpy, and kinetic fluxes). Most of the power transferred is emitted as auroral emission in the visible electromagnetic range, but a fraction is invested in cyclotron radio emission \cite{Zarka5} that escapes from the exoplanet environment if the surrounding stellar wind plasma frequency is smaller than the maximum cyclotron frequency at the planetary surface \cite{Weber,Vidotto4}. There are other sources of radio emission from giant gaseous exoplanets, where the acceleration of  electrons  is related to the rapid rotation of the magnetic field or its interaction with the plasma released by satellites or even their magnetosphere.

Radio telescopes lack the resolution of optical or infrared telescopes because the angular resolution, defined as $\lambda / D$ with $\lambda$ the observation wavelength and $D$ the telescope diameter, is smaller (the radio wavelength is $10^{5} - 10^{6}$ times larger than the visible wavelength). Consequently, the telescope diameter must be larger to reach the same angular resolution. To avoid this issue, current radio telescopes consist of arrays of wide spread antennas that can act as a single aperture, maximizing their collecting area and diameter. Array radio telescopes have been used to observe young stars and protoplanetary disks---for example, in 2014  the Atacama Large Millimeter Array (ALMA)  observed the young star HL Tau---finding gaps in the circumstellar disk identified as young planet-like bodies \cite{Ricci}.  The Very Large Array (VLA) also measured the radio emission protoplanetary disks in the star-forming region LDN 1551 \cite{Torrelles}. Radio receivers like the LOw Frequency ARray (LOFAR) operate in the frequency range of $10 - 240$ MHz with a sensitivity of $1$ mJy (at $15$ MHz) up to $0.3$ mJy (at $150$ MHz) \cite{Haarlem}. It should be noted that if the exoplanet magnetic field intensity is much lower than $4 \cdot 10^{5}$ nT the frequency of the signal is lower than $10$ MHz, below the LOFAR observation range. To be detectable from ground-based radio telescopes, the radio emission  needs to be in the range of $15$--$200$ MHz with the best chances for LOFAR if the emission is in the range of its peak sensitivity, between $50$ and $60$ MHz. In a recent study performed by LOFAR the radio emission from the Jovian radiation belts was measured \cite{Girard}. Another study performed at the Giant Meterwave Radio Telescope (GMRT) tentatively identified radio emission from a hot Jupiter \cite{Lecavelier}, but this result could not be reproduced, and is thus unconfirmed. By contrast, other attempts to detect the radio emission from exoplanet magnetospheres have failed because the telescope  was not sensitive enough \cite{Hallinan,Zarka6}, although the next generation of radio telescopes will be able to detect exoplanet radio emissions at  distances of $\le 20$ parsec \cite{Carilli,Nan,Ricci2}.  

The interaction of the stellar wind with the magnetosphere of an exoplanet can be described as the partial dissipation of the flow energy when a magnetized flow encounters an obstacle. The energy is dissipated as radiation in different ranges of the electromagnetic spectrum. This radiation depends on the flow and obstacle's  magnetic properties. The power dissipated ($[P_{d}]$) can be approximated as the intercepted flux of the magnetic energy ($[P_{d}] \approx B^2 v \pi R^{2}_{obs} / 2 \mu_{0}$), with $B$ the magnetic field intensity perpendicular to the flow velocity in the frame of the obstacle, $\mu_{o}$ the magnetic permeability of the vacuum, $v$ the flow velocity, and $R_{obs}$ the radius of the obstacle.

The radiometric Bode law links  the incident magnetized flow power and obstacle magnetic field intensity with the radio emission as $[P_{rad}] = \beta [P_{d}]^{n}$, with $[P_{rad}]$ the radio emission and $\beta$ the efficiency of dissipated power to radio emission conversion with $n \approx 1$ \cite{Zarka3,Zarka4}. Recent studies pointed out $\beta$ values between $2 \cdot 10^{-3}$ and $10^{-2}$ \cite{Zarka8}.

The interaction of the stellar wind (SW) with planetary magnetospheres is studied using different numerical frameworks such as single fluid codes \cite{2008Icar..195....1K,2015JGRA..120.4763J,Strugarek2,Strugarek}, multifluid codes \cite{2008JGRA..113.9223K}, and hydrid codes \cite{2010Icar..209...46W,Muller2011946,Muller2012666,2012JGRA..11710228R}. The simulations show that the planetary magnetic field is enhanced or weakened in distinct locations of the magnetosphere according to the IMF orientation, modifying its topology \cite{Slavin,2000Icar..143..397K,2009Sci...324..606S}. To perform this study we use the magnetohydrodynamic (MHD) version of the single fluid code PLUTO in spherical 3D coordinates \cite{Mignone}. The present study is based on previous theoretical studies devoted to simulating global structures of the Hermean magnetosphere using MHD numerical models \cite{Varela,Varela2,Varela3,Varela4}. The analysis takes part in recent modeling efforts to predict the radio emission from exoplanet magnetospheres \cite{Farrell2,Griemeier,Hess,Nichols}, complementary to other studies dedicated to analyzing the radio emission dependency with the stellar wind conditions \cite{Griemeier2,Jardine,Vidotto5,Vidotto6,See,Weber}. The model can be applied to exoplanet magnetospheres with topologies and intensities different from the Hermean case because no intrinsic spatial scales are contained in the equations of the ideal MHD (such as the Debye length or the Larmor radius in kinetic plasma physics) or in the spatial scale of the planetary dipole field. The only spatial scale of the problem is the planetary radius; however, this becomes negligible as soon as the magnetopause is far away from the planet surface (at least 2 times the planet radius). Under these circumstances, for the given stellar wind parameters  (i.e., sonic Mach number and plasma beta) there is no difference between a magnetosphere with standoff distance at $10$ planetary radii and a magnetosphere with standoff distance at  $1000$ planetary radii (and $10^{6}$ times stronger dipole field). The planet is essentially a point with no spatial scale in the simulation. Consequently, the study of the magnetospheric radio emission in exoplanets with a low or high magnetic field is analogous; from the point of view of the magnetosphere structure the problem to solve is the same. Foreseeing the magnetospheric radio emission in an exoplanet with a stronger magnetic field is a scaling problem that can be approximated to the first order using extrapolations. It should be noted that in a model with a strong magnetic field, any effects on the radio emission related to small magnetopause standoff distances are not observed, an important issue in exoplanets with large quadrupolar magnetic field components. Another reason to perform simulations with low magnetic fields is to maximize the model resolution required to reduce the numerical resistivity and obtain a better approach of the power dissipation in the magnetosphere. In addition, the inner boundary of the model is inside the exoplanet to reduce any numerical effects in the computational domain, so the Alfven time (the characteristic time of the simulation) should be small enough to have manageable simulations.

Previous studies predicted the variability of the power dissipated on the Hermean magnetosphere with the solar wind hydrodynamic parameters (density, velocity, and temperature) and interplanetary magnetic field orientation and intensity, and modified the topology of the Hermean magnetic field,   leading to different distributions of the energy dissipation hot spots (local maximum) and total emissivity \cite{refId0}. Exoplanet magnetospheres can also show very different configurations, for example a different ratio of dipolar to quadrupolar magnetic field components, magnetic axis tilt, intrinsic magnetic field intensity, rotation, distortions driven by the magnetic field of other planets or satellites,  leading to different exoplanet radio emissions even for the same configuration of the SW and IMF, namely host star of the same type, age, rotation, and magnetic activity.

The aim of this study is to estimate the radio emission driven in the interaction of the SW with an exoplanet magnetosphere, analyzing the kinetic and magnetic energy flux distributions as well as the net power dissipated on the exoplanet's day and night side, exploring the radio emission as a tool to identify the exoplanet's magnetic field properties. The analysis is performed for different orientations of the IMF and exoplanet magnetosphere topologies: different ratios of the dipolar to quadrupolar magnetic field components, magnetic axis tilts, and intrinsic magnetic field intensities. The parametrization of the radio emission in different exoplanet magnetosphere topologies is a valuable tool for the interpretation of future radio emission measurements and is used to estimate thresholds of the exoplanet magnetic field intensity, the ratio of the dipolar component versus higher degree components, or the magnetic axis tilt. Furthermore, if an exoplanet magnetosphere topology is known, the radio emission measurements also tabulate the instantaneous stellar wind dynamic pressure, as well as the IMF orientation and intensity of the host star at the exoplanet orbit \cite{Hess,Vidotto,Vidotto2}. 

This paper is structured as follows: Section 2, description of the simulation model, boundary and initial conditions; Section 3, analysis of the radio emission generation for exoplanets with different intrinsic magnetic field intensities; Section 4, analysis of the radio emission from an exoplanet magnetic field with different ratios of dipolar to quadrupolar components; Section 5, analysis of the radio emission for different tilts of the exoplanet magnetic axis; and Section 6, discussion and conclusions.

\section{Numerical model}

We use the ideal MHD version of the open source code PLUTO in spherical coordinates to simulate a single fluid polytropic plasma in the  nonresistive and inviscid limit \cite{Mignone}.

The conservative form of the equations are integrated using a Harten, Lax, Van Leer approximate Riemann solver (hll) associated with a diffusive limiter (minmod). The divergence of the magnetic field is ensured by a mixed hyperbolic--parabolic divergence cleaning technique \cite{Dedner}. The initial magnetic fields are divergenceless and remains so by applying the divergence cleaning method.

The grid is made of $256$ radial points, $48$ in the polar angle $\theta$ and $96$ in the azimuthal angle $\phi$ (the grid poles correspond to the magnetic poles). The simulation domain is confined within two spherical shells centered around the planet, representing the inner ($R_{in}$) and outer ($R_{out}$) boundaries of the system. Table 1 indicates the radial inner and outer boundaries of the system, characteristic length ($L$), effective numerical magnetic Reynolds number of the simulations due to the grid resolution ($R_{m}= V L/\eta$) and numerical magnetic diffusivity $\eta$. The kinetic Reynolds number ($R_{e}=V L/\nu$, with $\nu$ the numerical kinematic diffusivity) is in the range of the $[100,1000]$ for the different configurations. The numerical magnetic diffusivity is the driver of the reconnections in the model because we do not include an explicit value of the dissipation. Consequently, the numerical magnetic and kinetic diffusivities are determined by the grid resolution. \cite{Montgomery} calculated the kinematic viscosity and resistivity of the solar wind finding a value close to $1$ m$^2$/s and a Reynolds number of $10^{13}$. On the other hand, \cite{Verma} estimated an ion viscosity and resistivity of $5 \cdot 10^{7}$ m$^2$/s and a Reynolds number of $10^6$. If we assume \cite{Montgomery} results, the kinematic viscosity and resistivity values are too small and the Reynolds number too large to be simulated with the numerical resources available today because the grid resolution required is too large. If we consider the \cite{Verma} results, the numerical magnetic and kinetic diffusivities of the model for the given grid resolutions are closer to the solar wind parameters, particularly the B250 model (see columns VI and VII of table 1), so the study of the power dissipation should give a correct order of magnitude approximation. The numerical magnetic and kinematic diffusivity were evaluated in dedicated numerical experiments with the same grid resolution as the models and using a simpler setup, which indicated the limited impact of the grid resolution between models \cite{Varela,Varela2,Varela3,Varela4,Varela5,refId0}. The diffusivities change with the location because the grid is not uniform, so the dedicated experiments were performed using a resolution similar to the model resolution near the bow shock (BS) nose. It should be noted that the numerical resolution of the B6000 model is smaller than the B1000 and B250 models because the simulation domain is bigger for the same number of grid points, explaining why the numerical magnetic diffusivity is larger in that case.

Between the inner shell and the computational domain there is a ``soft coupling region'' ($R_{cr}$) where special conditions apply. Adding the soft coupling region improves the description of the plasmas flows towards the planet surface, isolating the simulation domain from spurious numerical effects of the inner boundary conditions \cite{Varela2,Varela3}. The outer boundary is divided into two regions, the upstream part where the stellar wind parameters are fixed and the downstream part where we consider the null derivative condition $\frac{\partial}{\partial r} = 0$ for all fields. At the inner boundary the value of the exoplanet intrinsic magnetic field is specified. In the soft coupling region the velocity is smoothly reduced to zero when approaching the inner boundary. The magnetic field and the velocity are parallel, and the density is adjusted to keep the Alfven velocity constant $\mathrm{v}_{A} = B / \sqrt{\mu_{0}\rho} = 25$ km/s with $\rho = nm_{p}$ the mass density, $n$ the particle number, and $m_{p}$ the proton mass. In the initial conditions we define a paraboloid on the night side with the vertex at the center of the planet, defined as $r < R_{cr} - 4sin(\theta)cos(\phi) / (sin^{2}(\theta)sin^{2}(\phi)+cos^{2}(\theta))$, where the velocity is null and the density is two orders of magnitude smaller than in the stellar wind. The IMF is also cut off at $R_{cr} + 2R_{ex}$, with $R_{ex}$ the exoplanet radius. Such initial conditions are required to stabilize the code at the beginning of the simulation. The radio emission of exoplanets with a magnetic field of $B_{ex}=6 \cdot 10^{3}$ nT is estimated on the day and night side using different $R_{out}$ values because several IMF orientations lead to the location of the magnetotail X point close to $R_{out}=150,$ while the magnetopause standoff distance is around $4R_{ex}$. Consequently, we use a model with $R_{out}=200$ to calculate the radio emission on the night side and a model with $R_{out}=75$ for the day side (improving the simulation resolution).

\begin{table*}[h]
\centering
\begin{tabular}{c | c c c c c c c}
Model & $R_{in}$ & $R_{out}$ & $R_{cr}$ & $L$ ($10^{6}$ m) & $R_{m}$ & $\eta$ ($10^{8}$ m$^{2}$/s) & $\nu$ ($10^{8}$ m$^{2}$/s)\\ \hline
B250 & $0.6$ & $16$ & $1.0$ & $2.44$ & $1800$ & $1.37$ & $0.30$\\
B1000 & $0.8$ & $30$ & $1.4$ & $4.30$ & $1020$ & $2.42$ & $0.42$\\
B6000 & $2.4$ & $75$ & $2.8$ & $36.59$ & $120$ & $25.81$ & $5.47$\\
Q02 & $2.0$ & $65$ & $2.5$ & $28.58$ & $150$ & $16.06$ & $4.10$\\
Q04 & $1.5$ & $50$ & $2.0$ & $17.82$ & $250$ & $10.01$ & $1.67$\\
\end{tabular}
\caption{Model inner boundary (column 1), outer boundary (column 2), soft coupling region (3), characteristic length (column 4), numerical magnetic Reynolds number (column 5), and numerical magnetic diffusivity (column 6).}
\label{table1}
\end{table*}

The exoplanet magnetic field is implemented in our setup as an axisymmetric ($m = 0$) multipolar field up to $l = 2$. The magnetic potential $\Psi$ is expanded in dipolar and quadrupolar terms:

\begin{equation} \label{eq:1}
\Psi (r,\theta) = R_{M}\sum^{2}_{l=1} (\frac{R_{M}}{r})^{l+1} g_{0l} P_{l}(cos\theta)  
\end{equation}

The current free magnetic field is $B_{M} = -\nabla \Psi $, $r$  the distance to the planet center, $\theta$ the polar angle, and $P_{l}(x)$ the Legendre polynomials. The numerical coefficients $g_{0l}$ for each model is summarized in Table 2. The model B6000 and the configurations with tilted magnetic axis have the same $g_{0l}$ coefficients. The effect of the tilt is emulated modifying the orientation of the IMF and stellar wind velocity vectors, so we can use the same setup of the axisymmetric multipolar field for all the models.

\begin{table}[h]
\centering
\begin{tabular}{c | c c }
Model & $g_{01}$(nT) & $g_{02}/g_{01}$   \\ \hline
B250 & $-250$ & $0$ \\
B1000 & $-1000$ & $0$ \\
B6000 & $-6000$ & $0$ \\
Q02 & $-4800$ & $0.25$ \\
Q04 & $-3600$ & $0.67$ \\
\end{tabular}
\caption{Multipolar coefficients $g_{0l}$ for the exoplanet internal field.}
\label{table2}
\end{table}

The simulation frame is such that the z-axis is given by the planetary magnetic axis pointing to the magnetic  north pole and the star is located in the XZ plane with $x_{star} > 0$. The y-axis completes the right-handed system. 

We assume a fully ionized proton electron plasma. The sound speed is defined as $\mathrm{v}_{s} = \sqrt {\gamma p/\rho} $ (with $p$ the total electron + proton pressure and $\gamma=5/3$ the adiabatic index), the sonic Mach number as $M_{s} = \mathrm{v}/\mathrm{v}_{s}$, and the Alfvenic Mach number as $M_{a} = \mathrm{v}/\mathrm{v}_{A}$. In the simulations the interaction of the stellar wind and the exoplanet magnetosphere leads to super-Alfvenic shocks ($M_{a}>1$), so the present model does not describe the radio emission from an exoplanet located in an orbit  where the interaction is sub-Alfvenic ($M_{a}<1$).

The recent model does not resolve the plasma depletion layer as a decoupled global structure from the magnetosheath due to the lack of model resolution, although simulations and observations share similar features in between the magnetosheath and magnetopause for the case of the Hermean magnetosphere \cite{Varela,Varela2,Varela3}. The magnetic diffusion of the model is larger than the real plasma so the reconnection between interplanetary and exoplanet magnetic field is instantaneous (no magnetic pile-up on the planet's day side) and stronger (enhanced erosion of the exoplanet magnetic field), although the essential role of the reconnection region in the depletion of the magnetosheath, injection of plasma into the inner magnetosphere and magnetosphere radio emission are reproduced \cite{refId0}. It should be noted that the exoplanet orbital motion is not included in the model, an effect likely important in the description of close-in exoplanets.

\section{Radio emission and exoplanet magnetosphere topology}

In  this section we estimate the radio emission of exoplanets with different magnetosphere topologies and IMF orientations. We calculate the power dissipated by the interaction of the SW with the exoplanet magnetosphere on the planet's day side and at the magnetotail X point on the planet's night side, leading to irreversible processes in which internal, bulk flow kinetic, magnetic, or system potential energy are transformed into accelerated electrons and then into radiation and heating sources on the exoplanet magnetosphere. The transfer of energy can be assessed by evaluating the various energy fluxes ($F$) involved in the system

\begin{equation}
\frac{\partial e}{\partial t} = -\vec{\nabla} \cdot \vec{F}
,\end{equation}
where

\begin{equation}
e = \rho \frac{v^{2}}{2} + \rho \frac{\gamma T}{\gamma - 1} + \frac{B^{2}}{2\mu_{0}}
,\end{equation}
and the energy flux

\begin{equation}
\vec{F} = \rho \vec{v} (\frac{v^{2}}{2}+\frac{\gamma T}{\gamma - 1}) + \vec{S} + \vec{Q}
.\end{equation}
The first term is the flux of kinetic energy, the second term is the enthalpy flux (the sum of internal energy and the potential of the gas to do work by expansion), the third term is the Poynting flux $\vec{S} = \vec{E} \wedge \vec{B} / \mu_{0}  \sim (\vec{v} \wedge \vec{B}) \wedge \vec{B}/\mu_{0}$ that shows the energy of the electromagnetic fields, and the last term $Q_{i} = -\mu \rho v_{i}\frac{\partial v_{i}}{\partial x_{j}}$ (i,j = 1,2,3) is the nonreversible energy flux. Here, $\vec{v}$ is the velocity field, $\vec{B}$ the magnetic field, $\vec{E}$ the electric field, $T$ the temperature, and $\mu$ the shear viscosity.

In the following, we calculate the power dissipated as a combination of the kinetic $P_{k}$ (associated with the stellar wind dynamic pressure) and magnetic Poynting $P_{B}$ terms (due to the reconnection between the IMF and the exoplanet magnetic field). The enthalpy and the nonreversible energy flux are neglected because they are tiny. The net power dissipated is calculated as the volume integral of $P_{k}$ and $P_{B}$ divergence in the regions of energy dissipation associated with hot spots (we define the threshold at $|P_{B}| > |P_{B}|_{max}/3$, with $|P_{B}|_{max}$ the absolute value of the maximum magnetic power in the hot spots):

\begin{equation} \label{eq:4}
[P_{k}] = \int_{V}  \vec{\nabla} \cdot \left(\frac{\rho \vec{\mathrm{v}} |\vec{\mathrm{v}}^{2}|}{2} \right) dV 
\end{equation}
\begin{equation} \label{eq:5}
[P_{B}] = -\int_{V} \vec{\nabla} \cdot \frac{(\vec{\mathrm{v}}\wedge\vec{B})\wedge\vec{B}}{\mu_{0}} dV
\end{equation}

On the day side, the volume integrated extends from the BS to the inner magnetosphere near the radio emission hot spots (magnetosheath and magnetopause included). On the night side the integrated volume is localized in the magnetotail X point where the magnetic field module is smaller than $10$ nT.

\subsection{Interaction between IMF and exoplanet magnetic field}

We  now show examples of the interaction between the interplanetary and the exoplanet magnetic fields. In the following, the hydrodynamic parameters of the stellar wind in the simulations are fixed, summarized in Table 3 (including the stellar wind plasma frequency $\omega_{sw}$), as is the module of the IMF that is kept to $20$ nT (the IMF orientation for each model can be found in the Table B.1 of the Appendix). The selected IMF module and SW dynamic pressure are the expected typical values for an exoplanet in an orbit near  the habitable zone of a star similar to the Sun (between 0.95 and 1.67 au; \cite{Kopparapu,Gallet}). We only consider the typical values because the SW and IMF instantaneous parameters can be very variable, for example in the case of Mercury, the range of SW density possible values is $[10 , 180]$ cm$^{-3}$, velocities between $[300 , 700]$ km/s and temperatures of [45,000, 200,000] K \cite{Varela2}. For a systematic study on the effect of the stellar wind dynamic pressure, temperature, and IMF on the radio emission we refer the reader to \cite{refId0}.

\begin{table}[h]
\centering
\begin{tabular}{c c c c c c c c}
$n$ (cm$^{-3}$) & $T$ (K) & $\mathrm{v}$ (km/s) & $\omega_{sw}$ (kHz) \\ \hline
$60$ & $90000$ & $350$ & $69.5$\\
\end{tabular}
\caption{Hydrodynamic parameters of the SW}
\label{table3}
\end{table}

Figure 1 shows a 3D view of the system for a northward IMF orientation of a model with a dipolar magnetic field of $6000$ nT and a magnetic axis tilt of $60^{o}$ (with respect to the rotation axis). The BS is identified by the color scale of the density distribution (large density increase). The SW dynamic pressure bends the exoplanet magnet field lines (red lines) compressing the magnetic field lines on the exoplanet's day side and forming the magnetotail on the night side. The IMF (pink lines) reconnects with the exoplanet magnetic field lines leading to the formation of the exoplanet magnetopause. The arrows indicate the orientation of the IMF and the dashed white line the outer limit of the simulation domain (the star is not included in the model).

\begin{figure}[h]
\centering
\resizebox{\hsize}{!}{\includegraphics[width=\columnwidth]{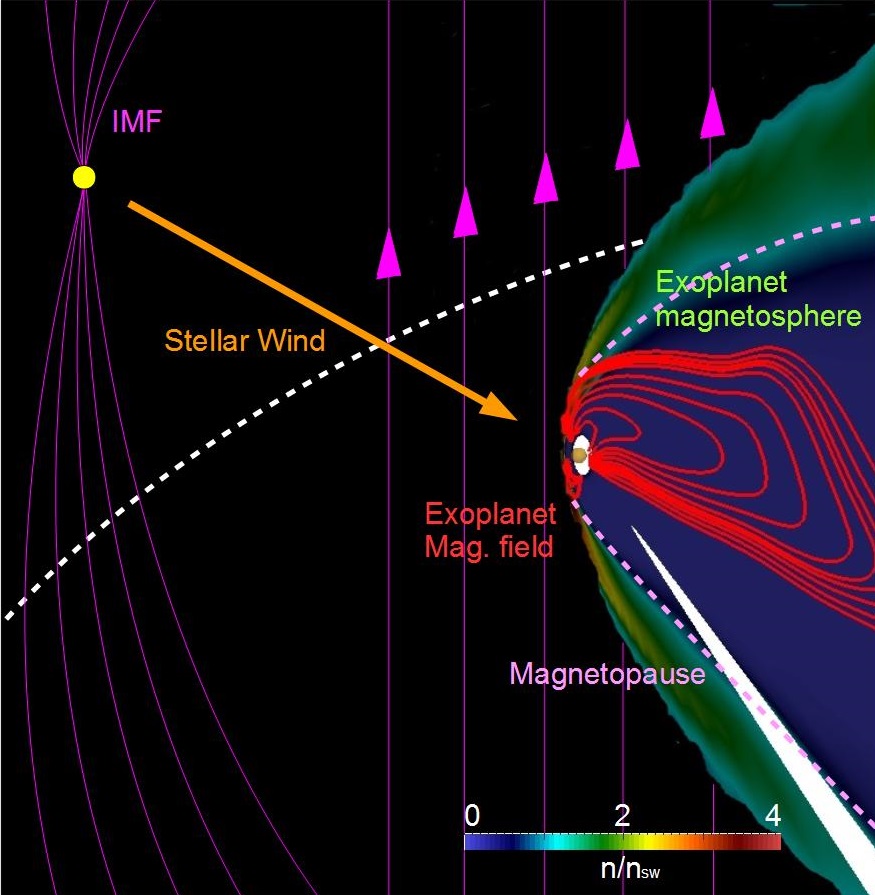}}
\caption{Three-dimensional view of a typical simulation setup. Density distribution (color scale), exoplanet magnetic field lines (red lines), and IMF (pink lines). The arrows indicate the orientation of the IMF (northward orientation). The dashed white line shows the beginning of the simulation domain.  We note that the star is not included in the model.}
\label{1}
\end{figure}

In the following we identify the IMF orientation from the exoplanet to the star as Bx simulations, the IMF orientation from the star to the exoplanet as Bxneg simulations, the northward orientation with respect to the exoplanet's magnetic axis as Bz simulations (figure 1), the southward orientation as Bzneg simulations, the orientation perpendicular to the previous two cases on the exoplanet orbital plane as By (east) and Byneg (west) simulations. The IMF simulations  and exoplanet intrinsic magnetic configurations are summarized in  Appendix B.

\begin{figure}[h]
\centering
\resizebox{\hsize}{!}{\includegraphics[width=\columnwidth]{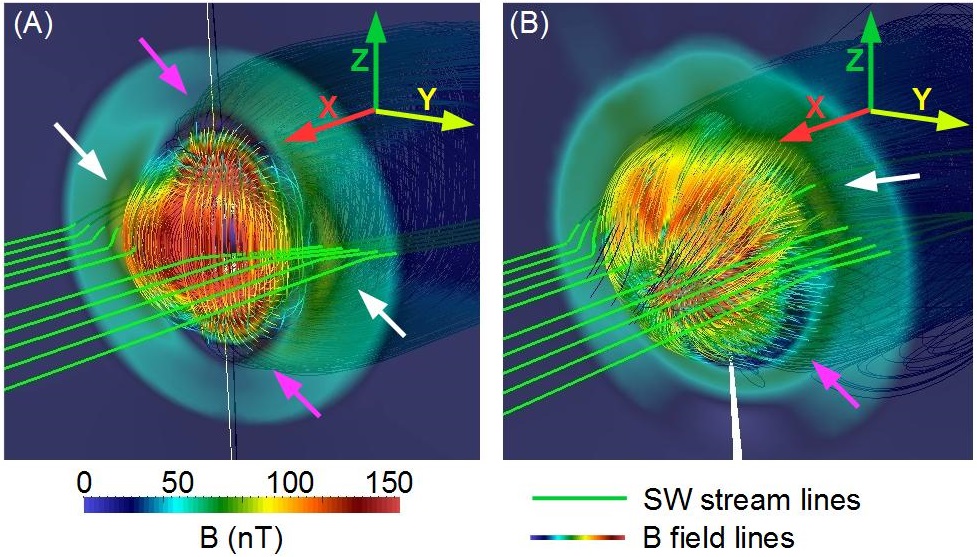}}
\caption{Exoplanet magnetic field lines with the intensity imprinted on the field lines by a color scale for the B6000 configuration with a magnetic axis tilt of $0^{o}$ (A) and $60^{o}$ (B). Magnetic field intensity at the frontal plane $X = 3R_{ex}$. Stellar wind stream lines (green). The pink arrow shows the reconnection region and the white arrow the magnetic field pile up region. The IMF is oriented in the Bx direction}.
\label{2}
\end{figure}

Figure 2 illustrates the interaction of the IMF and the exoplanet magnetospheric field in the model B6000 (panel A) and model tilt60 (panel B). The exoplanet magnetic field lines are color-coded with the magnetic field amplitude and the green lines are the SW stream lines. The frontal plane at $X = 3R_{ex}$ shows the magnetic field intensity. The reconnection regions are identified as a local decrease (blue color near the poles for model B6000 and near the south pole for model tilt60, highlighted by pink arrows) and local magnetic field pile up as an increase (yellow/orange colors near the equator for model B6000 and near the north pole for the model tilt60, highlighted by white arrows) of the magnetic field. The reconnections are associated with regions of SW injection in the inner magnetosphere (plasma streams from the magnetosheath towards the exoplanet surface) \cite{Varela3}. The magnetic field pile up regions are linked with radio emission hot spots (acceleration of electrons along the magnetic field lines) \cite{refId0}. In both cases there is a conversion of magnetic energy into kinetic and internal energy. Consequently, the exoplanet magnetic field topology and IMF orientation are critical in addressing  the exoplanet radio emission since it is the direct outcome of the location and intensity of the magnetosphere reconnection and magnetic field pile up regions. It should be noted that the energy dissipation and radio emission hot spots are not localized in the same regions of the magnetosphere; the electrons are accelerated in the zones with large energy dissipation whereas the radio emission is generated along their trajectory around the magnetic field lines towards the exoplanet surface where the cyclotron frequency is the highest. Nevertheless, the energy dissipation and radio emission hot spots are correlated and show some common features in the simulations.

The present study is limited to the analysis of the dissipated power and radio emission driven by the stellar wind interaction with the magnetosphere of rocky and giant gaseous exoplanets. It should be noted that the radio emission from giant gaseous exoplanets is also caused by internal plasma sources combined with their fast rotation, as  was observed for Jupiter and to a lesser extent for Saturn \cite{Bagenal,Krupp}, conditions not included in our present model so the predicted values may be considered as a lower bound. Icy exoplanets similar to Uranus or Neptune show strongly nonaxisymmetric magnetic fields and fast rotation, so an axisymmetric magnetic field model cannot  reproduce their radio emission properly.

\subsection{Effect of the exoplanet magnetic field intensity}

In this section we analyze the effect of the intrinsic magnetic field intensity on the radio emission for different IMF orientations. We perform simulations for exoplanets with an intrinsic magnetic field intensity of $250$, $1000,$ and $6000$ nT.

\begin{figure*}[h]
\centering
\resizebox{\hsize}{!}{\includegraphics[width=\columnwidth]{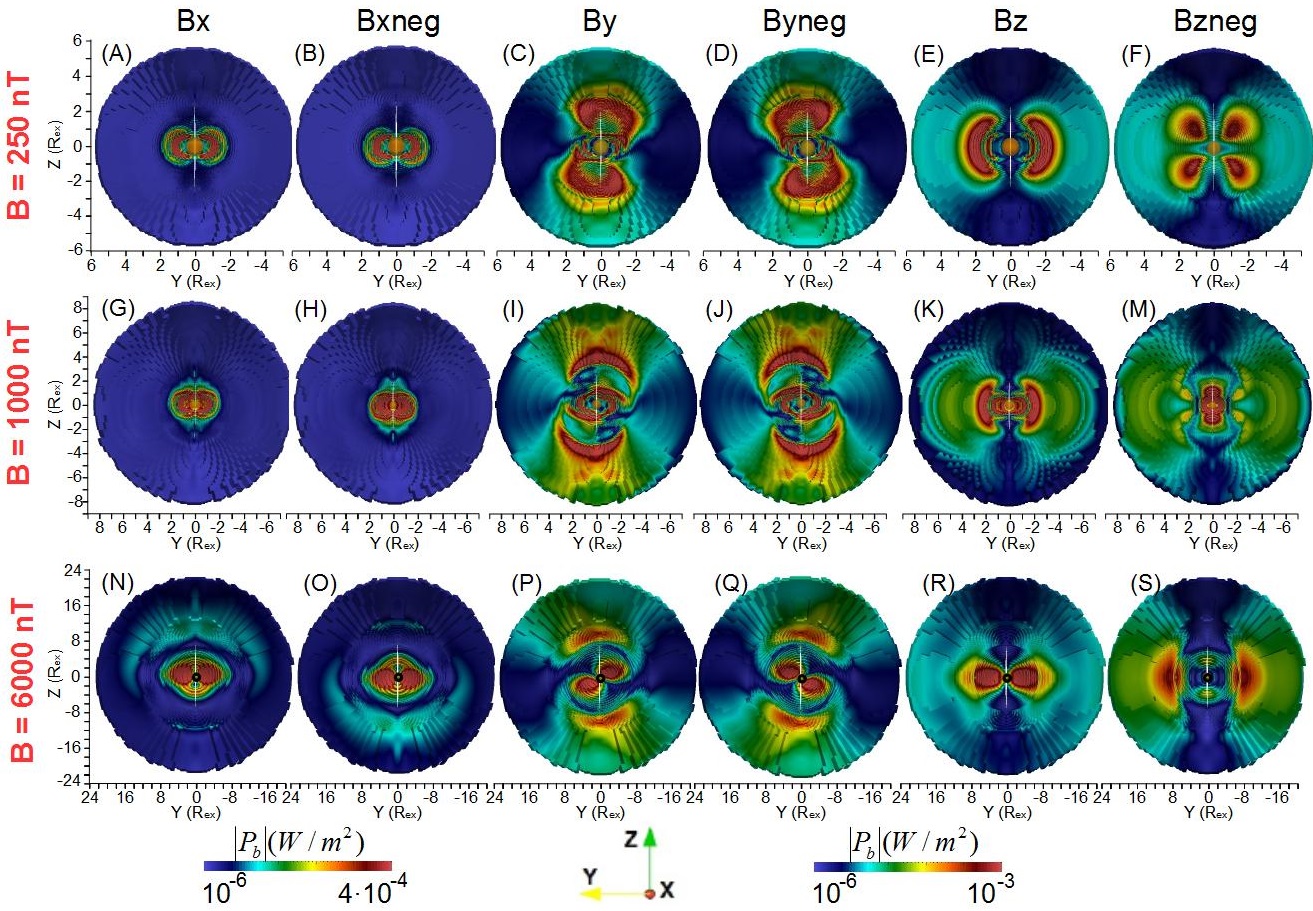}}
\caption{View of the magnetic power from the night side of the exoplanet for different IMF orientations and exoplanet intrinsic magnetic field intensities. The first color bar is related to the Bx--Bxneg IMF orientations, and the second color bar is related to the other IMF orientation. The plotted surface is defined between the bow shock and the magnetopause where the magnetic power reaches its maxima.}
\label{3}
\end{figure*}

Figure 3 shows a view of the magnetic power from the night side of the exoplanet ($P_{B}(DS)$) for different IMF orientations and exoplanet intrinsic magnetic field intensities. The minima of the magnetic power are correlated with a local decrease in the exoplanet magnetic field intensity, while the maxima are correlated with a local enhancement of the exoplanet magnetic field (pile up). The hot spot distribution for the Bx--Bxneg IMF orientations is north--south asymmetric (panels 3A, B, G, H, N, and O) because there is reconnection region at the south (north) of the magnetosphere if the IMF is oriented in the Bx (Bxneg) direction. The hot spots are displaced northward for Bx IMF orientations and southward for Bxneg IMF orientations as the exoplanet magnetic field intensity increases because the reconnection regions are located  farther away from the exoplanet surface (the magnetopause standoff distance increases). The hot spot distribution for the  By--Byneg IMF orientations shows an east--west asymmetry also correlated with the location of the reconnection regions (panels 3C, D, I, J, P, and Q). If the exoplanet magnetic field intensity increases the hot spots sideways the magnetic axis are located farther away from the exoplanet, caused by the increase in the magnetopause standoff distance and the counterclockwise (co) rotation of the hot spots for the By (Byneg) IMF orientation. The reconnection regions for the Bz (panels E, K, and R) and Bzneg (panels 3F, M, and S) IMF orientations are located near the exoplanet poles and the equator, respectively. If the exoplanet magnetic field intensity increases the reconnection regions are located farther away from the poles (equatorial region), and the hot spot distribution is  located closer to the equatorial (polar) region. In summary, the hot spots are located farther away from the exoplanet surface as the magnetic field intensity increases. It should be noted that a larger SW dynamic pressure leads to a more compact magnetosphere on the exoplanet's day side, so the hot spots will be located closer to the exoplanet surface. Consequently, the correct identification of the exoplanet magnetic field intensity requires an accurate identification of the host start SW dynamic pressure at the exoplanet orbit (a deviation larger than a $50 \%$ from the real value, particularly if the stellar wind pressure is large, will lead to incorrect results). Such information can be partially inferred analyzing the radio emission from the exoplanet's night side  because a strong radio emission is a sign of intense magnetotail stretching and high SW dynamic pressure \cite{Varela2}.

\begin{figure*}[h]
\centering
\resizebox{\hsize}{!}{\includegraphics[width=\columnwidth]{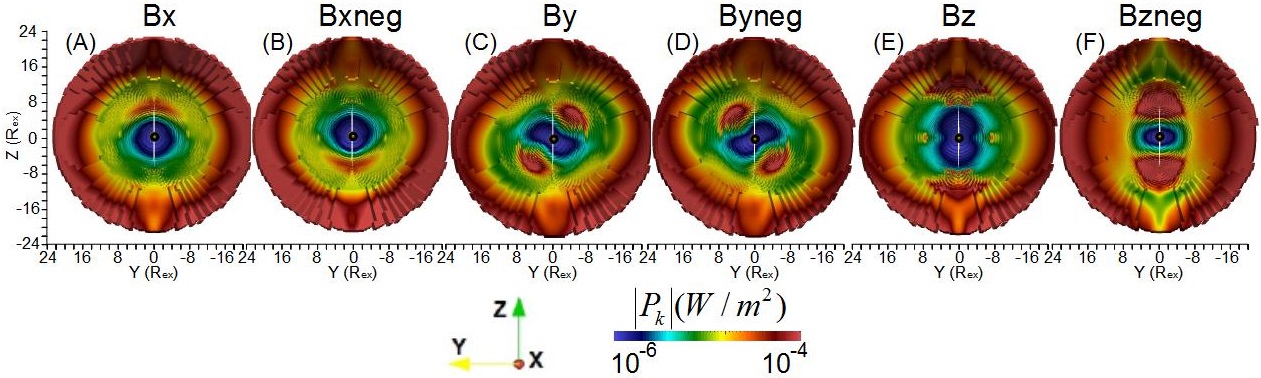}}
\caption{View of the kinetic power from the night side of the exoplanet in the B6000 model for different IMF orientations. The plotted surface is defined between the bow shock and the magnetopause where the kinetic power reaches its maxima.}
\label{4}
\end{figure*}

Figure 4 shows a view of the kinetic power from the night side of the exoplanet ($P_{k}(DS)$) of B6000 model for different IMF orientations. A local decrease (enhancement) in the magnetospheric field is associated with a local enhancement (decrease) in $P_{k}(DS)$ caused by the acceleration of the plasma in the reconnection regions where the stellar wind is injected in the inner magnetosphere. Consequently, the $P_{k}(DS)$ distribution is determined by the magnetosphere topology and IMF orientation. An increase of the SW dynamic pressure enhances the magnetic and kinetic powers, although the hot spot distribution is only slightly disturbed \cite{refId0}. 

\begin{figure*}[h]
\centering
\resizebox{\hsize}{!}{\includegraphics[width=\columnwidth]{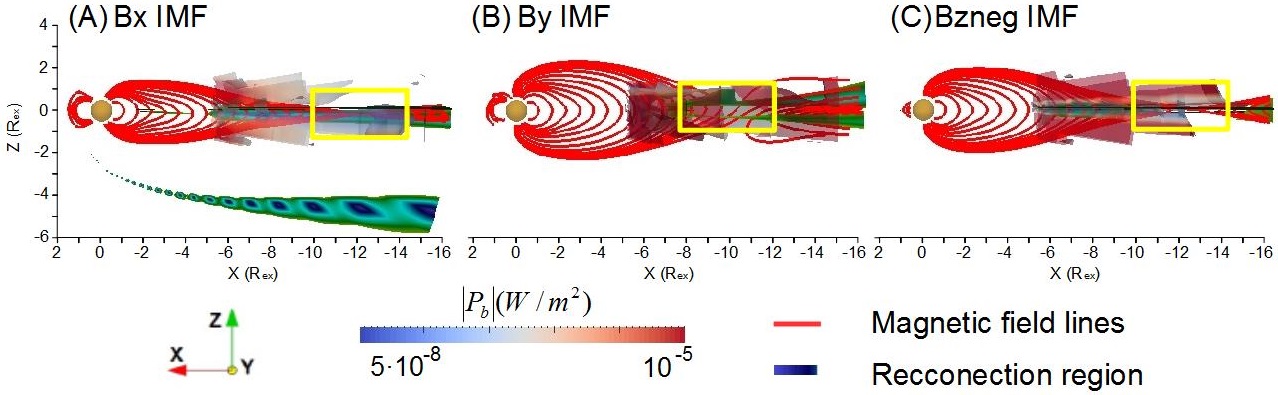}}
\caption{Magnetic power on the exoplanet's night side (PB(NS)) for model B250 for Bx (A), By (B), and Bnegz (C) IMF orientations. The reconnection region (isosurface with magnetic field intensity between 0 and 20 nT) is indicated in dark blue and dark green;   the magnetotail reconnection region is indicated by the yellow rectangle. Magnetic field lines of the exoplanet and IMF are indicated in red.}
\label{5}
\end{figure*}

Figure 5 shows a zoomed view of the magnetic power ($P_{B}(NS)$) on the night side for an exoplanet with a magnetic field intensity of $250$ nT for Bx, By, and Bzneg IMF orientations. The different IMF orientations modify the exoplanet's magnetosphere topology (red magnetic field lines) and reconnection regions between the exoplanet magnetic field and the IMF (dark blue and dark green iso-surfaces linked to the magnetopause and the magnetotail X point). For this reason the location and intensity of the radio emission hot spots are different in each model. The last closed magnetic field line indicates the location of the magnetopause and the open magnetic field lines are reconnected lines between the IMF and the exoplanet magnetic field.

The  expected radio emission is calculated from the net magnetic and kinetic power dissipated on the planet's day and night sides using the radiometric Bode law \cite{Zarka3,Farrell} for different exoplanet magnetic field intensities:
\begin{equation} \label{eq:6}
[P(DS)] = a [P_{k}(DS)] + b [P_{B}(DS)]
\end{equation}
\begin{equation} \label{eq:7} 
[P(NS)] = a [P_{k}(NS)] + b [P_{B}(NS)]
\end{equation} 
with $a$ and $b$ the efficiency ratios assuming a linear dependency of $[P_{k}]$ and $[P_{B}]$ with $[P]$. The radio emission measured from the solar system planets can be explained by two possible combinations of efficiency ratios: ($a = 10^{-5}$, $b=0$) or ($a = 0$, $b=2\cdot10^{-3}$) \cite{Zarka4,Zarka10}. In the following, we only consider the combination of parameters $a = 0$, $b=2\cdot10^{-3}$ because the other combination leads to a radio emission several orders of magnitude smaller on the day and night sides. All the $[P(DS)]$ and $[P(NS)]$ values are calculated for an exoplanet with the same radius as Mercury ($R_{ex}= 2440$) km to have a straightforward comparison with the \cite{refId0} results. The model is in adimensional units and the distance is normalized to the exoplanet radius, so $[P(DS)]$ and $[P(NS)]$ can be expressed in terms of any exoplanet radius considering that $[W] = kg m^{2} / s^{3}$. For example, if we calculate the radio emission of an exoplanet with the same radius as the Earth, the values in the tables must be multiplied by a factor $(R_{Earth}/R_{Mercury})^2= 6.67$. It should be noted that the radius of the obstacle in the analysis of the radio emission is the distance from the magnetopause to the exoplanet surface, not the exoplanet radius;  the radio power is enhanced as the module of the exoplanet magnetic field increases because the magnetosphere is wider. On the other hand the ratio between the exoplanet radius must be included in the extrapolation to be consistent with the fact that the magnetic field module is compared at the exoplanet surface.

\begin{table}[h]
\centering
\begin{tabular}{c}
$[P(DS)]$ ($10^{5}$ W)
\end{tabular}

\begin{tabular}{c | c c c c c c}
Model & Bx & Bxneg & By & Byneg & Bz & Bzneg \\ \hline
B250 & $0.63$ & $0.63$ & $6.90$ & $8.72$ & $5.96$ & $11.3$ \\
B1000 & $2.18$ & $2.04$ & $12.4$ & $12.2$ & $10.6$ & $29.5$ \\
B6000 & $4.34$ & $4.21$ & $35.0$ & $32.5$ & $14.7$ & $72.5$ \\
\end{tabular}

\begin{tabular}{c}
Linear regression slope day side vs $B_{ex}$ ($10^{11}$ W/T)
\end{tabular}

\begin{tabular}{c | c c c c c c}
 & Bx & Bxneg & By & Byneg & Bz & Bzneg \\ \hline
$\alpha$ & $0.77 $ & $0.74$ & $6.05$ & $5.65$ & $2.71$ & $12.6$ \\
$\Delta \alpha$ & $\pm 0.1$ & $\pm 0.1$ & $\pm 1$ & $\pm 1$ & $\pm 1$ & $\pm 2$ \\
\end{tabular}

\begin{tabular}{c}
$[P(NS)]$ ($10^{5}$ W) 
\end{tabular}

\begin{tabular}{c | c c c c c c}
Model & Bx & Bxneg & By & Byneg & Bz & Bzneg \\ \hline
B250 & $0.10$ & $0.10$ & $0.53$ & $0.53$ & $0.13$ & $0.18$ \\
B1000 & $1.04$ & $0.72$ & $1.74$ & $1.71$ & $1.00$ & $0.66$ \\
B6000 & $10.2$ & $10.2$ & $15.8$ & $16.4$ & $7.75$ & $20.4$ \\
\end{tabular}

\begin{tabular}{c}
Linear regression slope night side ($10^{11}$ W/T)
\end{tabular}

\begin{tabular}{c | c c c c c c}
 & Bx & Bxneg & By & Byneg & Bz & Bzneg \\ \hline
$\alpha$ & $1.68$ & $1.67$ & $2.61$ & $2.70$ & $1.28$ & $3.32$ \\
$\Delta \alpha$ & $\pm 0.08$ & $\pm 0.1$ & $\pm 0.1$ & $\pm 0.1$ & $\pm 0.04$ & $\pm 0.3$ \\
\end{tabular}

\label{table4}
\caption{Expected radio emission on the exoplanet's day side (first table) and in the magnetotail X point on the exoplanet's night side (third table) for different IMF orientations ($a = 0$, $b=2\cdot10^{-3}$) and exoplanet magnetic field intensities. Slope ($\alpha$) and goodness of fit ($\Delta \alpha$) of the linear regression $[P] = \alpha B_{ex}$ for each IMF orientation (second and fourth table). $B_{ex}$ is the magnetic field intensity on the exoplanet surface and the exoplanet radius is $R_{ex}=2440$ km.}
\end{table}

Table 4 shows the predicted radio emission on the exoplanet's day side (top rows) and night side (middle rows). The radio emission increases with the magnetic field intensity, consistent with the theoretical scaling confirmed by the radio emission measurements of the gaseous planets in the solar system \cite{Desch,Zarka2,Zarka3,Zarka4,Nichols}. It should be noted that the scaling law ``emitted power'' versus ``impinging Poynting flux'' only gives an order of magnitude estimation for what may be observable with a given radio telescope, so the current paper is not about detection but about emission efficiency for various exoplanet magnetic field configurations and stellar wind magnetic field orientations. In addition, the total radio power is integrated over all the frequencies of the emission although a radio telescope has a finite bandwidth, so the radio power in Table 4 overestimates the radio telescope measurements. The radio emission on the day side is almost one order of magnitude higher than the radio emission on the night side for all the IMF orientations in models B250 and B1000. On the other hand the radio emission is larger on the night side in the model B6000 for the Bx--Bxneg IMF orientations. If we calculate the linear regression $[P] = \alpha B_{ex}$ (third table), with $B_{ex}$ the magnetic field intensity on the exoplanet surface, we observe that only the IMF orientations Bx--Bxneg lead to a stronger radio emission on the night side. For the By--Byneg and Bz--Bzneg IMF orientations the radio emission on the day side is 2 times larger than the night side. The fit goodness of the linear regression ($\Delta \alpha$) shows a reasonable agreement with the data tendency. The radio emission on the night side of model B6000 shows a smaller variation between the different IMF orientations, because the internal magnetosphere topology is less affected by the IMF orientation as the exoplanet magnetic field intensity increases. The strongest radio emission on the day side is observed for the Bzneg IMF orientation, followed by the By--Byneg orientations, whereas the Bx--Bxneg IMF orientations lead to the weakest radio emission. Thus, future radio emission measurements require an observation time, long enough, to remove the effect of the IMF orientation (as well as the variation in the SW dynamic pressure) because the instantaneous radio emission can change by up to one order of magnitude if the IMF is oriented in the exoplanet--star, southward or northward orientations. Similar trends are reproduced in a previous study about the IMF effect on the Hermean magnetospheric radio emission \cite{refId0}.

The maximum cyclotron frequency at the planetary surface for the models B250, B1000, and B6000 is $f_{max} = 14$, $56,$ and $336$ kHz. Consequently, the radio emission from exoplanets with less than $B_{ex} \approx 1000$ nT is less likely to be observed (at least for the stellar wind conditions analyzed in this article, where the plasma frequency is $69.5$ kHz). Based on our knowledge at Jupiter, Saturn, and the Earth, the auroral  radio emission (CMI) is produced between very low frequencies (kHz) and $f_{max}$. It should be noted that the radio-magnetic scaling law relates the total power integrated on the emission’s spectrum and beaming pattern (and average time variations). In addition, the average radio flux density can be deduced from the power divided by the spectral range ($\approx f_{max}$) and the solid angle in which the emission is beamed (typically $0.2$ to $1$ sr, see \cite{Zarka9}). The peak instantaneous flux density can exceed the average flux density by $2$ orders of magnitude \cite{Galopeau,Zarka9,Lamy}.

\subsection{Effect of the magnetic field quadrupolar-to-dipolar components ratio}

In this section we analyze the effect of the exoplanet magnetic field topology on the radio emission in configurations with different ratios of the dipolar and quadrupolar components, namely models Q02 ($B_{dip}= 0.8 \cdot B_{ex}$ and $B_{quad}= 0.2 \cdot B_{ex}$ nT) and Q04 ($B_{dip}= 0.6 \cdot B_{ex}$ and $B_{quad}= 0.4 \cdot B_{ex}$ nT) with $B_{ex}=6000$ nT, for different IMF orientations.

Figure 6 shows a polar cut of the density distribution (color scale, panels A and B) and the frontal plane of the magnetic field modulus (color scale, panels C and D) of the models Q02 and Q04 for a Bx IMF orientation. The red lines show the exoplanet magnetic field lines. Compared to the B6000 model, Q02 and Q04 modes show wider regions of open magnetic field lines on the planet surface and a smaller magnetopause standoff distance because  an increase in the $g_{20}/g_{10}$ ratio leads to the northward displacement and a faster decay of the exoplanet magnetic field. Consequently, the magnetosphere topology of the models Q02 and Q04 is different regarding the model B6000 so the reconnection regions, dissipation, and radio emission hot spot locations and intensity also change. Moreover, a higher $g_{20}/g_{10}$ ratio leads to a thinner and deformed magnetosheath, so the SW precipitates directly towards the exoplanet surface at low southern hemisphere latitudes in the Q4 model.

\begin{figure}[h]
\centering
\resizebox{\hsize}{!}{\includegraphics[width=\columnwidth]{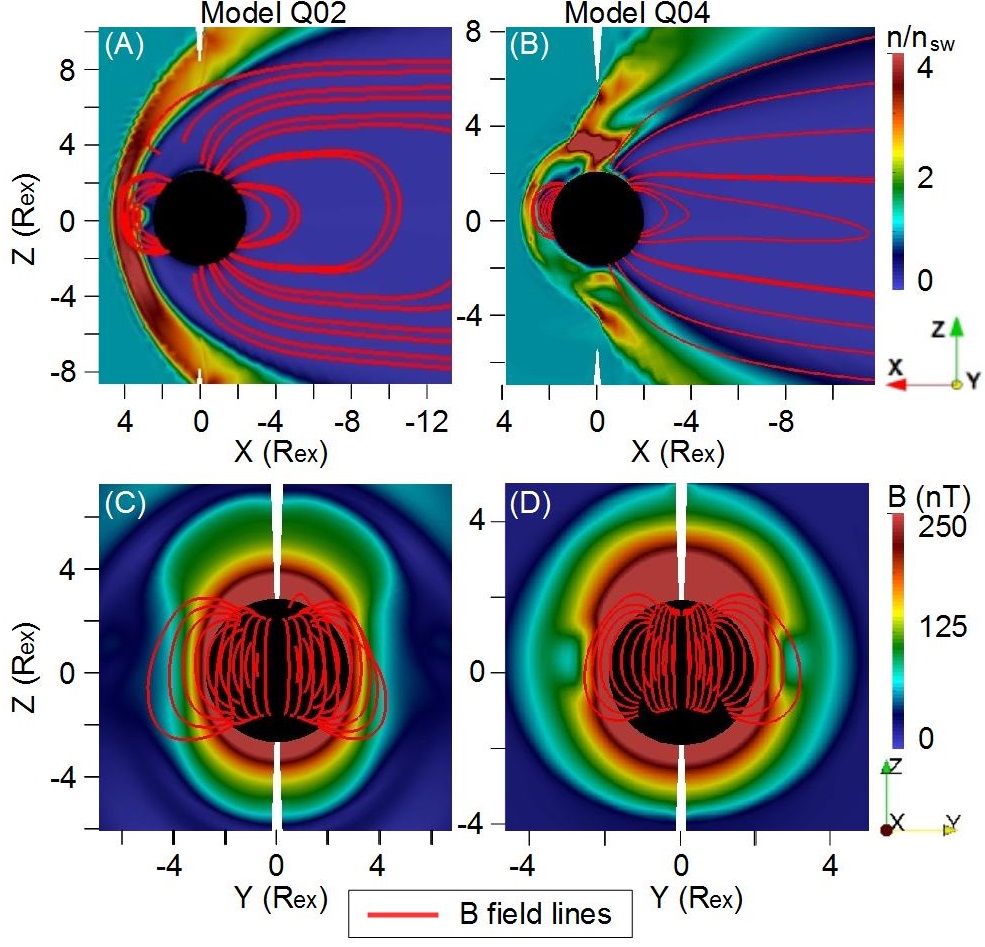}}
\caption{Polar cut of the density distribution (color scale) and field lines of the exoplanet magnetic field (red lines) of models Q02 (A) and Q04 (B) for the Bx IMF orientation. The black disk radius is $R_{cr}$. Frontal cut of the magnetic field module on the star--exoplanet direction of models Q02 (C) and Q04 (D).}
\label{6}
\end{figure}

Figure 7 shows a view of the magnetic power from the night side of the exoplanet for different IMF orientations and exoplanet magnetic field topologies. If the IMF is oriented in the  Bx--Bxneg directions (panels A, B, G, and H), an increment of the quadrupolar component of the exoplanet magnetic field leads to a northward drift of the hot spots, located farther away from the north pole for the Bx IMF orientation and closer to the equator for the Bxneg IMF orientation regarding the B6000 model, due to the northward displacement of the magnetosphere. A similar effect is observed for the By--Byneg IMF orientations (panels C, D, I and J) where the hot spots in the north of the magnetosphere are located farther away from the exoplanet, although the hot spots in the south of the magnetosphere are located closer to the exoplanet. For the Bz--Bzneg IMF orientations the hot spots are also displaced northward. The Q04 model shows for all the IMF orientations a region of large magnetic power near the exoplanet (panels K and M);  compared to the B6000 and Q02 models (panels E and F) the magnetopause standoff distance is smaller and the reconnection regions are enhanced, which is caused by the strong deformation of the internal magnetospheric field compared to the dipolar case. These results point out that a northward displacement (or southward depending on the exoplanet magnetic field orientation)  of the hot spot distribution, independently of the instantaneous IMF orientation, indicates a large quadrupolar component of the exoplanet magnetic field. It should be noted that the radio telescope observation angle with respect to the exoplanet and the exoplanet--host star vector must be calculated accurately to avoid an overestimation of the quadrupolar component.

\begin{figure*}[h]
\centering
\resizebox{\hsize}{!}{\includegraphics[width=\columnwidth]{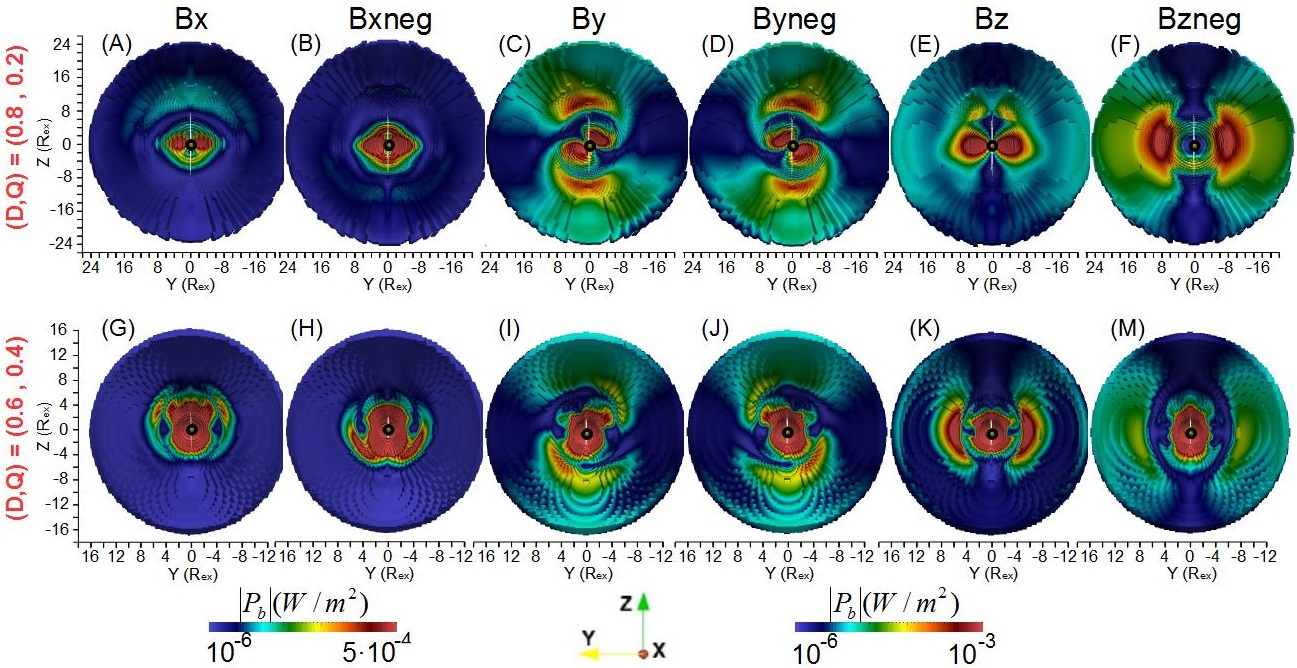}}
\caption{View of the magnetic power from the night side of the exoplanet for different IMF orientations and exoplanet magnetic field topologies (different dipolar-to-quadrupolar component ratios). The first color bar is related to the Bx--Bxneg IMF orientations, and the second color bar is related to the other IMF orientations. The plotted surface is defined between the bow shock and the magnetopause where the magnetic power reaches its maxima.}
\label{7}
\end{figure*}

Figure 8 shows the magnetic power ($P_{B}(NS)$) on the exoplanet's night side and the magnetosphere topology of the model Q04 for the Bx, By, and Bz IMF orientations. Compared to the B6000 model (see figure 5) the magnetotail is slender and stretched, a consequence of a stronger perturbation of the internal magnetosphere topology by the IMF due to a faster decay of the quadrupolar component with respect to the dipolar component, so the radio emission is lower. To quantify the magnetotail stretching we calculate the ratio between the averaged width of the magnetotail with the location of the X point, showing values around $0.115$ for the Bx IMF case, $0.169$ for the By IMF case, and $0.187$ for the Bz case, smaller than the B6000 model where the ratio for the Bx IMF case is $0.313$ ($2.7$ times larger), for the By IMF case is $0.399$ ($2.3$ times larger), and for the Bz IMF case is $0.355$ ($1.9$ times larger). Consequently, the radio emission on the exoplanet's night side varies almost by one order of magnitude between the different configurations.

\begin{figure*}[h]
\centering
\resizebox{\hsize}{!}{\includegraphics[width=\columnwidth]{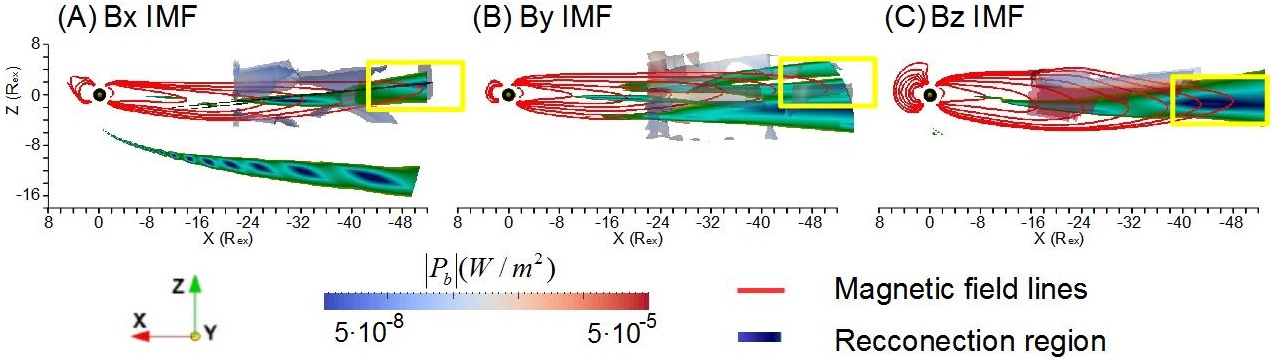}}
\caption{Magnetic power on the exoplanet's  night side (PB(NS)) of model Q04. The reconnection region (isosurface with magnetic field intensity between 0 and 20 nT) is indicated in dark blue and dark green;  the magnetotail reconnection region is indicated by the yellow rectangle. Magnetic field lines of the exoplanet and IMF are indicated in red.}
\label{8}
\end{figure*}

Table 5 shows the expected radio emission for different ratios of the quadrupolar to dipolar components on the exoplanet's day side (the values on the night side are not shown because the trends only indicate a decrease in the radio emission as the quadrupolar-to-dipolar ratio increases due to the faster decay of the quadrupolar component):

\begin{table}[h]
\centering
\begin{tabular}{c}
$[P(DS)]$ ($10^{5}$ W)
\end{tabular}

\begin{tabular}{c | c c c c c c}
Model & Bx & Bxneg & By & Byneg & Bz & Bzneg \\ \hline
Q02 & $3.33$ & $4.40$ & $19.1$ & $16.9$ & $9.43$ & $33.5$ \\
Q04 & $26.6$ & $51.3$ & $63.9$ & $55.6$ & $42.1$ & $44.6$ \\
\end{tabular}

\begin{tabular}{c}
Model / B6000 $[P(DS)]$ ratio
\end{tabular}

\begin{tabular}{c | c c c c c c}
Model & Bx & Bxneg & By & Byneg & Bz & Bzneg \\ \hline
Q02 & $0.77$ & $1.05$ & $0.55$ & $0.52$ & $0.64$ & $0.46$ \\
Q04 & $6.13$ & $12.2$ & $1.83$ & $1.71$ & $2.86$ & $0.61$ \\
\end{tabular}

\label{table5}
\caption{Expected radio emission on the exoplanet's day side for different IMF orientations ($a = 0$, $b=2\cdot10^{-3}$) and exoplanet magnetic field intensities. Results for an exoplanet with a radius of $R_{ex}=2440$ km.}
\end{table}

The radio emission of the model Q02 on the day side decreases for all the IMF orientations (except for the Bxneg case, which shows a slight increase) because the faster decay of the quadrupolar component leads to a weaker reconnection region showing similar internal magnetosphere topology than the B6000 model. If the quadrupolar component is large enough, as in  model Q04, the internal magnetospheric topology changes with respect to the B6000 model leading to an enhancement of the reconnection regions and the radio emission near the exoplanet surface.

\subsection{Effect of the exoplanet magnetic axis tilt}

In this section we analyze the effect of the magnetic axis tilt on the exoplanet radio emission, namely the models tilt30 (tilt=$30^{o}$), tilt60 (tilt=$60^{o}$), and tilt90 (tilt=$90^{o}$). The analysis of the magnetic axis tilt is performed in addition to the study of the IMF orientation because different angles between magnetic axis and stellar wind velocity vector leads to different exoplanet  magnetosphere configurations, due to the effect of the stellar wind dynamic pressure.

Figure 9 shows a view of the magnetic power from the night side of the exoplanet for different IMF orientations and exoplanet magnetic axis tilts. The hot spots  distribution for the Bx (panels A and G) and Bxneg (panels B and H) IMF orientations are displaced respectively southward and northward  as the tilt increases from $0^{o}$ to $60^{o}$ because the reconnection regions are displaced closer to (or farther away from) the star and closer to the exoplanet equatorial plane. In addition, models with a large tilt and a Bx (Bxneg) IMF orientation show a hot spot distribution similar to models with small tilt and a Bz (Bzneg) IMF orientation (compare panels N and O of fig. 4 with panels K, M, R, and S of fig. 9, or panels R and S of fig. 4 with panels G and H of fig. 9). This comes about because the magnetosphere topology is almost analogous if the SW dynamic pressure is not strong enough to drive major deformations on the magnetosheath structure. Compared to the B6000 model the hot spots are more spread out and located farther from the exoplanet by the effect of the SW dynamic pressure because as the tilt increases the IMF is more aligned with the SW flow. The model tilt90 has a reconnection region in the exoplanet equatorial plane where the IMF and magnetic field lines are (anti-)parallel if the IMF is oriented in the Bx (Bxneg) direction. Thus, the SW precipitates directly towards the surface at the equator and there is an enhancement of the magnetic power near the exoplanet. The hot spot distribution is located in the region with closed magnetic field lines, forming a ring around the exoplanet in the XY plane. For the By--Byneg IMF orientations, the hot spots at the north of the magnetosphere are located farther away from the exoplanet and more aligned with the magnetic axis as the magnetic axis tilt increases, so the east--west asymmetry of the magnetosphere decreases (panels C, D, I, J, P, and Q).

\begin{figure*}[h]
\centering
\resizebox{\hsize}{!}{\includegraphics[width=\columnwidth]{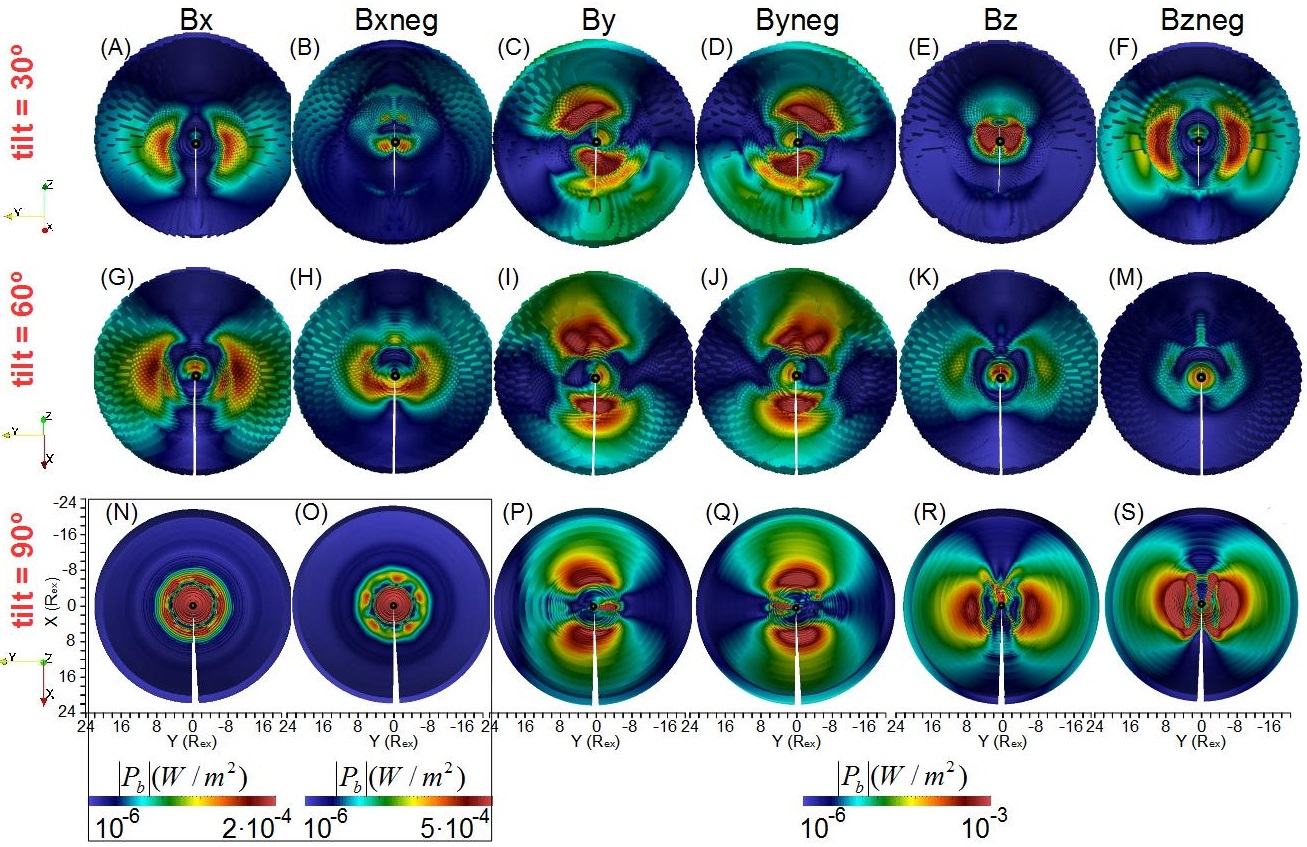}}
\caption{View of the magnetic power from the night side of the exoplanet for different IMF orientations and exoplanet magnetic axis tilts. The color bars of the model tilt90 for the Bx--Bxneg IMF orientations are different from those in  the rest of the panels. The plotted surface is defined between the bow shock and the magnetopause where the magnetic power reaches its maxima.}
\label{9}
\end{figure*}

Figure 10 shows the magnetic power ($P_{B}(NS)$) of the models tilt30, tilt60, and tilt90 on the planet's night side for the Bx IMF orientation. The magnetotail topology changes as the tilt increases, showing a more slender and stretched shape, so the radio emission also changes for each configuration. The ratio between the magnetotail width and X point location is $0.232$ for the Bx IMF tilt30 case,  $0.212$ for the Bx IMF tilt60 case, and $0.135$ for the Bx IMF tilt90 case. Model tilt90 is an extreme case with a reconnection ring in the YZ plane due to the bending of the closed magnetic field lines by the SW at the north and south geographic poles towards the star--exoplanet direction. 

\begin{figure*}[h]
\centering
\resizebox{\hsize}{!}{\includegraphics[width=\columnwidth]{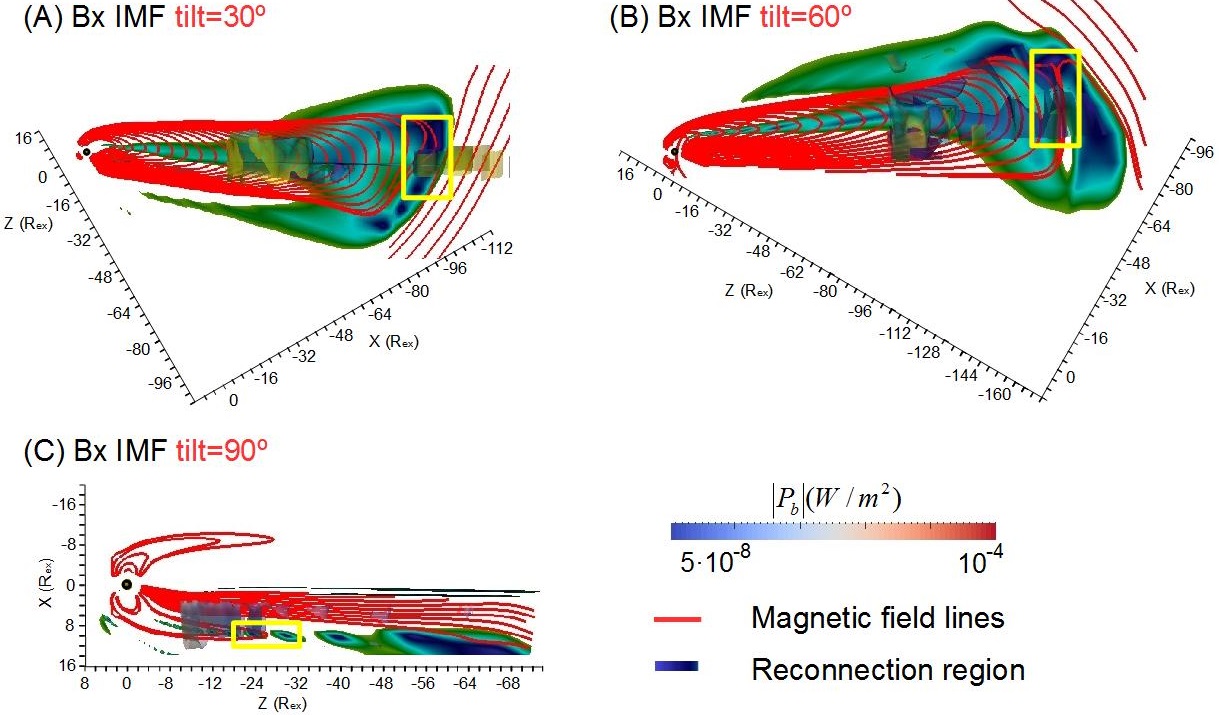}}
\caption{Magnetic power on the exoplanet's night side (PB(NS)) of models tilt30, tilt60, and tilt90 for the Bx IMF orientation. Panels (G) and (H) show  models tilt30 and tilt90 for a Bx IMF orientation. The reconnection region (isosurface with magnetic field intensity between 0 and 20 nT) is indicated in dark blue and  dark green; the magnetotail reconnection region is indicated by the yellow rectangle. Magnetic field lines of the exoplanet and IMF are indicated in red.}
\label{10}
\end{figure*}

Table 6 shows the radio emission for different magnetic axis tilts on the planet's day side. Increasing the magnetic axis tilt from $0^{o}$ to $60^{o}$ leads to a stronger radio emission for the Bx--Bxneg IMF orientations because the reconnection regions and hot spots are wider. On the other hand, if the tilt is $90^{o}$ the radio emission for the Bx (Bxneg) IMF orientation decreases (increases) due to the increase (decrease) in the exoplanet magnetic field near the magnetic poles by the effect of the magnetic reconnections (see fig. 9). If the IMF is oriented in the By--Byneg IMF direction, the radio emission increases between $0^{o}$ to $30^{o}$ because the hot spots are wider, although from $30^{o}$ to $90^{o}$ both radio emission and hot spot size saturate. For the Bz--Bzneg IMF orientations, the radio emission and hot spot size decreases as the tilt increases. On the other hand, the radio emission and hot spot size increases in the model tilt90. On the exoplanet's night side, the radio emission increases with the tilt because the magnetotail stretching is greater.

\begin{table}[h]
\centering
\begin{tabular}{c}
$[P(DS)]$ ($10^{5}$ W)
\end{tabular}

\begin{tabular}{c | c c c c c c}
Model & Bx & Bxneg & By & Byneg & Bz & Bzneg \\ \hline
tilt30 & $55.1$ & $19.6$ & $87.3$ & $80.1$ & $7.36$ & $49.9$ \\
tilt60 & $74.6$ & $32.4$ & $81.5$ & $82.0$ & $2.94$ & $4.06$ \\
tilt90 & $18.7$ & $57.6$ & $10.8$ & $7.87$ & $19.1$ & $60.1$ \\
\end{tabular}

\begin{tabular}{c}
Model / B6000 $[P(DS)]$ ratio
\end{tabular}
\begin{tabular}{c | c c c c c c}
Model & Bx & Bxneg & By & Byneg & Bz & Bzneg \\ \hline
tilt30 & $12.7$ & $6.66$ & $2.49$ & $2.46$ & $0.5$ & $0.69$ \\
tilt60 & $17.2$ & $7.70$ & $2.33$ & $2.52$ & $0.20$ & $0.06$ \\
tilt90 & $4.34$ & $13.7$ & $0.31$ & $0.24$ & $1.30$ & $0.83$ \\
\end{tabular}

\begin{tabular}{c}
$[P(NS)]$ ($10^{5}$ W) 
\end{tabular}

\begin{tabular}{c | c c c c c c}
Model & Bx & Bxneg & By & Byneg & Bz & Bzneg \\ \hline
tilt30 & $70.7$ & $72.8$ & $95$ & $108$ & $48.9$ & $108$ \\
tilt60 & $98.3$ & $89.8$ & $114$ & $119$ & $62.1$ & $192$ \\
\end{tabular}

\begin{tabular}{c}
Model / B6000 $[P(NS)]$ ratio
\end{tabular}

\begin{tabular}{c | c c c c c c}
Model & Bx & Bxneg & By & Byneg & Bz & Bzneg \\ \hline
tilt30 & $6.93$ & $7.14$ & $6.01$ & $6.59$ & $6.31$ & $5.29$ \\
tilt60 & $9.64$ & $8.80$ & $7.21$ & $7.26$ & $8.01$ & $9.41$ \\
\end{tabular}

\label{table6}
\caption{Expected radio emission on the exoplanet's day and night sides for different IMF orientations ($a = 0$, $b=2\cdot10^{-3}$) and magnetic axis tilts. Results for an exoplanet with a radius of $R_{ex}=2440$ km.}
\end{table}

\section{Discussion and conclusions}
\label{Conclusions}

The aim and main contribution of the present communication is to show the radio emission as a potential tool for identifying the exoplanet's magnetic field properties. The information provided will be useful to guide future radio emission measurements to infer the exoplanet's magnetosphere properties such as the magnetic field intensity, tilt angle, and topology for different IMF orientations.

The analysis shows that the energy dissipation, hot spot distribution, and total radio emission are correlated with the exoplanet magnetic field topology and IMF orientation. Different magnetospheric configurations lead to different locations of the reconnection regions and hot spot distributions on the exoplanet's day side, associated with the maximum of the magnetic power and the minimum of the kinetic power, as well as local enhancements of the magnetospheric field. Therefore the characteristics of the exoplanet's magnetic field could likely be inferred by future radio telescopes because the radio emission measurements contain information about the exoplanet's magnetic field intensity, dipolar-to-quadrupolar components ratio, and magnetic axis tilt. The present and planned low-frequency radio telescopes with high sensitivity will reach $0.1$'' to $1$'' angular resolution, enough to separate an exoplanet from a star if the exoplanet orbits at several AU and the system is not farther than a few tens of parsecs, but will unlikely resolve structures within the exoplanetary magnetosphere. On the other hand, from the modeling of the dynamic spectrum in intensity and circular polarization it is possible to deduce several physical parameters from the system, including the planet’s magnetic field amplitude, tilt, offset, planetary rotation period, or the inclination of the orbital plane \cite{Hess}.

An increase in the exoplanet magnetic field intensity leads to an enhancement of the radio emission on the exoplanet's day and night sides, and to an increase in the hot spot size located farther away from the exoplanet surface (Fig. 11A). A linear regression between the exoplanet magnetic field and the radio emission on the day and night sides also shows a reasonable agreement. In addition, a large quadrupolar component of the exoplanet magnetic field leads to a northward (or southward) displacement of the magnetospheric field and the hot spot distribution. If the quadrupolar component is large enough ($g_{20}/g_{10} > 2/3$), the internal magnetospheric field is deformed compared to a pure dipolar case, leading to a larger radio emission on the exoplanet's day side (Fig. 11, panel C), although the radio emission on the night side decreases due to the faster decay of the quadrupolar component compared to the dipolar component.

The models with a small tilt and a Bx (Bxneg) IMF orientation have a similar magnetospheric topology to  configurations with large tilt and Bz (Bzneg) IMF orientation (and vice versa), although not the same because the angle between the magnetic axis and stellar wind velocity vector is different. The magnetosphere topology can be different if the stellar wind dynamic pressure is large enough to drive strong deformations on the magnetosheath, for example if the host star has strong stellar wind fluxes or the exoplanet is in an orbit close to the host star. The consequence is an enhancement of the radio emission on the day side as the tilt increases for the Bx--Bxneg IMF orientations and a decrease for the Bz--Bzneg IMF orientations (Fig 11D). On the other hand if the exoplanet magnetic axis tilt is large the hot spot distribution is more spread out due to the effect of the SW dynamic pressure, because the SW is more aligned with the magnetic poles leading to an increase in the radio emission.

The radio emission on the night side is dominant for the Bx--Bxneg IMF orientations if the exoplanet magnetic field intensity is higher than $1000$ nT, although the radio emission on the day side is larger for the other IMF orientations (Fig. 11, panel B). If we extrapolate the trends obtained for the radio emission and exoplanet magnetic field intensity on the day and night sides (using the stellar wind parameters listed in Table 3), the expected radio emission range of a hot Jupiter with $B_{ex}=5 \cdot 10^{5}$ nT and a radius of $R_{ex}=7.2 \cdot 10^4$ km (similar to Jupiter) is $[P(DS)] = 0.3 - 5.5 \cdot 10^{11}$ W and $[P(NS)] = 0.7 - 1.4 \cdot 10^{11}$ W and of a super Earth with $B_{ex}=6 \cdot 10^{4}$ nT and $R_{ex}=1.26 \cdot 10^4$ km is $[P(DS)] = 0.6 - 9.6 \cdot 10^{8}$ W and $[P(NS)] = 1.3 - 2.5 \cdot 10^{8}$ W, values consistent with the observational scaling \cite{Desch2,Zarka7,Zarka3,Nichols}. Previous numerical studies predicted the radio emission of a hot Jupiter located $3$ to $10$ radius away from a star similar to the Sun \cite{Nichols}, suggesting a value of $[5,1300] 10^{12}$ W for an exoplanet magnetic fields between $0.1$ to $10$ times  Jupiter's magnetic field, several orders of magnitude above the present model expectations. The reason for this difference is the dynamic pressure, almost $3\cdot10^{3}$ times lower in the present model. As a proxy of the magnetic power enhancement with the increase in the dynamic pressure we consider the results of \cite{refId0}:  a dynamic pressure $3000$ times higher leads to a radio emission enhancement of 2700 times with respect to the present model.  This means that  the expected maximum magnetic power of the model is $[P_{B}(DS)] \approx 1.5 \cdot 10^{18}$ W and $[P_{B}(NS)] \approx 0.4 \cdot 10^{18}$ W, similar to the analysis performed by \cite{Saur} and \cite{Strugarek},  who expected a magnetic power around $10^{19}$ W. For the same reason the observational scaling shows a radio emission value almost one order of magnitude higher than the Jupiter radio emission measurements because the dynamic pressure at the Jupiter orbit is lower. Nevertheless, the real radio emission must be larger in a hot Jupiter with respect to the present results because we do not add the effect of other radio emission sources such as the fast rotation or internal plasma releases that do not depend on the distance to the parent star, thus the extrapolation result may be considered as a lower bound. If the hot Jupiter is located at $20$ parsec, the radio emission flux at Earth can be calculated as $\Phi = P / \Omega d^2 \omega$ with $\Omega \approx 1.6$ sr the solid angle, $d$ the distance to the exoplanet, and $\omega = 15$ MHz the detection bandwidth of the receiver, leading to a value of $0.1 - 1$ mJy, in the limit of the LOFAR observation range. For the case of a super Earth $\Phi \leq 10^{-4}$ mJy. The radio emission flux is lower than $10^{-4}$ mJy in the simulations performed in our study (e.g.,  model B6000 with a Bzneg IMF orientation, $\Phi = 3 \cdot 10^{-5}$ mJy), although stronger wind conditions lead to a higher radio emission flux. In addition, the model is only representative of an exoplanet with dipolar field without magnetic axis tilt in an orbit similar to Mercury for a host star like the Sun. If the exoplanet is located closer to the host star the SW dynamic pressure and IMF intensity is higher so the radio emission is also enhanced. Likewise, if the host star magnetic activity is higher the IMF is also larger (stars younger than the Sun with faster rotation are more active; see, e.g., \cite{Emeriau}) as well as the radio emission. The SW and IMF characteristics also change if the host star is not the same type as the Sun, leading to a different scaling \cite{Reville}. In other words, dedicated analyses are required to foresee the threshold of the exoplanet magnetic field topology for each star spectral type, age, magnetic activity, and exoplanet orbital distance \cite{Jardine,Vidotto5,See,Vidotto6,Weber}.

The radio emission in models with different IMF orientation show a variability factor up to 20, describing how the radio emission of an exoplanet should change during the magnetic cycle of the host star or through IMF variations along the exoplanet orbit \cite{Vidotto3}. Such variability can partially explain the uncertainty in the determination of the average auroral radio powers using the radio Bode law, around one order of magnitude between the lower and upper bounds \cite{Zarka3}.

The radio emission can escape the exoplanet magnetosphere if the maximum CMI emission is larger  than the plasma frequency in the surrounding stellar wind.  In the case of Mercury, the maximum CMI emission probably does not exceed a few 10s kHz, whereas the plasma frequency in the surrounding solar wind is between $70 - 100$ kHz, thus the CMI radiation---if it exists (which should be confirmed by BepiColombo and MMO measurements)---is trapped in the magnetospheric cavity. For close-in exoplanets, if the planet’s exo-ionosphere is expanding the CMI emission can be trapped inside the magnetosphere \cite{Weber}, although there are several possibilities for overcoming the CMI emission trapping; for example, if the exoplanet shows small-scale auroral plasma cavities like at the Earth, there are second harmonic emissions (especially on the ordinary mode) or if the exoplanet magnetic field is strong. Another option is to analyze the radio emission from exoplanets located farther away from the host star where the plasma frequency is lower.

Using the results of the  present study  it is possible to identify, in a first approximation, the minimum exoplanet radio emission associated with a magnetic field strong enough to shield the exoplanet surface (at low latitudes) from the stellar wind. The exoplanet magnetopause standoff distance can be estimated by this simplified expression:
$$ \frac{R_{MP}}{R_{ex}} = \left(\frac{B_{ex}^{2}}{m_{p}n\mu_{0}\mathrm{v}^{2}}\right)^{1/6} $$
Here $R_{MP}$ is the magnetopause standoff distance and $m_{p}$ the proton mass (we consider the same SW dynamic pressure as in the simulations, see Table 3). If the ratio is $ R_{MP} / R_{ex} = 1$ the SW precipitates directly toward the exoplanet surface, so the magnetic field is not strong enough to shield the planet, namely $B_{ex} \approx 120$ nT. If the exoplanet has the same radius as the Earth and the magnetic field is a dipole without magnetic axis tilt, the radio emission range is $[P(DS)] = 0.6 - 10 \cdot 10^{5}$ W and $[P(NS)] = 1.3 - 2.7 \cdot 10^{5}$ W, so we can identify a threshold for the exoplanet habitability from the point of view of the radio emission output: if the radio emission measurement is lower than $[P] = 10^{6}$ W the exoplanet is less likely to host life on the surface \cite{Vidotto7,Vidotto8}. There are other restrictions for the exoplanet habitability, for example the host star age. If the star is younger than the Sun the magnetic activity is higher, due to a faster rotation, so extreme events such as the coronal mass ejections (CME) are more frequent \cite{Aarnio}, which is  why the exoplanet habitability requires a stronger magnetic field with a larger magnetopause standoff distance \cite{Khodachenko,Lammer}. If  $ R_{MP} / R_{ex} = 5$, the exoplanet surface will be also shielded from most of the CME, so the exoplanet magnetic field must be at least $B_{ex} \approx 1.5 \cdot 10^{4}$ nT, leading to a radio emission of $[P(DS)] = 0.7 - 13 \cdot 10^{7}$ W and $[P(NS)] = 1.7 - 3.3 \cdot 10^{7}$ W. Thus, we can define another radio emission threshold for exoplanet habitability  of $[P] = 1.3 \cdot 10^{8}$ W if the host star is younger and more
active than the Sun (but the same star type). Compared to the case of the Earth (older host star with lower SW dynamic pressure and IMF intensity at the exoplanet orbit), previous studies indicate a radio emission of at least $10^{7}$ W for a magnetosphere that can host life on the surface \cite{Zarka5}, a value compatible with the present results. 

It should be noted that the expression to calculate the standoff distance only provides an approximated value, because a dipolar magnetic field with no tilt is assumed and the effect of the IMF orientation is not considered, so dedicated numerical experiments are required to obtain more accurate thresholds. In addition, this results are only valid if the SW dynamic pressure and IMF intensity at the exoplanet orbit are similar to the case of Mercury.  

The combination of efficiency ratios ($a = 0$, $b=2\cdot10^{-3}$) shows the highest radio emission values. A previous study of the radio emission from the Hermean magnetosphere identified these efficiency ratios as the most accurate option for reproducing the expected radio emission from Mercury \cite{refId0}, but these results should be confirmed by in situ measurement and radio emission data from gaseous planets of the solar system. The present study's results can also be  used to compare the expect radio emission of Bode's law with radio telescope measurements, with the aim of inferring the efficiency ratios that most closely fit the observations for different SW dynamic pressures, IMF orientations, and planet magnetic field topologies.

The trends obtained in the analysis are useful for the exoplanet magnetospheres detectable by the current radio telescopes, particularly hot Jupiters. Among other conclusions, we show that the present radio telescopes have enough sensitivity to measure the hot Jupiter radio emission, and   in the best cases can possibly even constrain their magnetic field topology. In addition, the model can be calibrated analyzing the radio emission from the gaseous planets of the solar system---the results of the analysis  and the scaling are similar  to these measurements and to models of  other authors---and in the near future by the measurement of the radio emission from Mercury by the BEPIColombo satellite.

The net magnetic energy dissipation predicted by  Bode's law and the simulations show  good agreement, so the magnetic energy dissipation on the day side and the magnetotail reconnection regions are well reproduced by the model in a first approximation. On the other hand, the net kinetic energy dissipation predicted by the simulations is smaller than calculated by Bode's law because the model can reproduce more accurately the complex flows on the day and night sides of the exoplanet.

\begin{figure}[h]
\centering
\resizebox{\hsize}{!}{\includegraphics[width=\columnwidth]{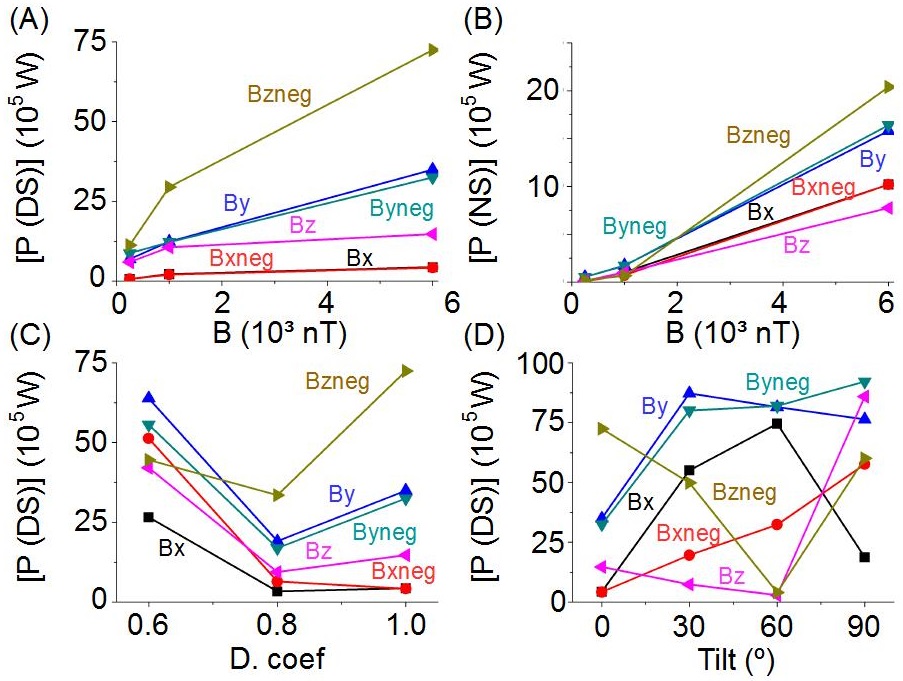}}
\caption{(A) Radio emission on the day side vs exoplanet magnetic field intensity. (B) Radio emission on the night side vs the exoplanet magnetic field intensity. (C) Radio emission on the day side vs the ratio of the exoplanet quadrupolar to dipolar magnetic field components. (D) Radio emission on the day side vs the magnetic axis tilt. Results for an exoplanet with a radius of $R_{ex}=2440$ km.}
\label{11}
\end{figure}

Future measurements of the radio emission will allow  testing of the different configurations of the exoplanet magnetospheres in the light of the model selection problem in Bayesian statistics \cite{William}. In other words, it will be possible to select the  model that best reproduces the observations based on the computation of the Bayesian evidence \cite{Trotta,Corsaro1,Corsaro2}, a key parameter that provides a statistical weight, favoring models that provide a better fit to the data but penalizing those that have a more complex analytical representation, i.e., a larger number of parameters that configure the model itself. In this way it will be possible to unambiguously constrain the most favored theoretical interpretation for a given observational set.

In summary, radio emission data bring constraints on the exoplanet magnetosphere topology,  essential information to foresee the potential habitability of exoplanets, associated with the presence of  permanent and strong enough magnetic fields to shield the planet surface and atmosphere from the stellar wind erosion. This information can be deduced if we analyze large time series of radio emission data when available \cite{Hess,Zarka6}, removing the effect of the instantaneous effect of the IMF orientation, intensity, and  the stellar wind dynamic pressure and temperature. On the other hand, after identifying the characteristics of the exoplanet magnetic field, the radio emission data is useful in order to determine the properties of the stellar wind and magnetic field of the star. This process can be carried out through the adoption of a Bayesian model comparison. In this view, the competing models to test will incorporate the different configurations of the exoplanet magnetospheres, as shown in this work, and will be fit to the radio emission data in order to obtain the best set of free parameters which best match the observed radio emission. In a subsequent step, the Bayesian evidence of each model are compared in order to select the most likely configuration that reproduces the same observational set \cite{Corsaro2}. Our aim is to develop this thorough statistical approach by computing a grid of predictive models for future releases of radio emission measurements, and to  test the methodology using synthetic datasets. The final target is to create a catalog that illustrates the main features of the exoplanets' magnetic fields and identify those that can harbor life.

\ack
This material is based on work supported  by both the U.S. Department of Energy and the Office of Science, under Contract DE-AC05-00OR22725 with UT-Battelle, LLC. The research leading to these results has received funding from the European Commission's Seventh Framework Programme (FP7/2007-2013) under the grant agreement SHOCK (project number 284515), the grant agreement SPACEINN (project number 312844), and ERC STARS2 (207430). We extend our thanks to CNES for Solar Orbiter and PLATO science support and to INSU/PNST for our grant. We acknowledge GENCI allocation 1623 for access to the supercomputer where most of the simulations were run, and DIM-ACAV for supporting the ANAIS project and our graphics/post-analysis and storage servers, as well as DIO of the Paris Observatory. The authors would also like to acknowledge R.A. Garcia and E. Corsaro for the fruitful discussion.

This manuscript has been authored by UT-Battelle, LLC under Contract No. DE-AC05- 00OR22725 with the U.S. Department of Energy. The United States Government retains and the publisher, by accepting the article for publication, acknowledges that the United States Government retains a non exclusive, paid-up, irrevocable, world wide license to publish or reproduce the published form of this manuscript, or allow others to do so, for United States Government purposes. The Department of Energy will provide public access to these results of federally sponsored research in accordance with the DOE Public Access Plan (http://energy.gov/downloads/doe-public-access-plan).

\begin{appendix}

\section{Calculation of the numerical magnetic diffusivity}
 
We performed a set of simulations in a simplified test case to analyze the numerical magnetic diffusivity in a downscale model of characteristic length $L=1$ m with a grid made of $196$ radial points, $48$ in the polar angle and $96$ in the azimuthal angle for $R_{out}=12$. We study the evolution of a 3D Gaussian profile in a motionless fluid:
$$ \frac{\partial \vec{B}}{\partial t} + \cancel{\vec{\nabla} \wedge (-\vec{v} \wedge \vec{B})} = -\eta \frac{1}{\mu_{0}} \vec{\nabla} \wedge (\vec{\nabla} \wedge \vec{B})$$
$$ \Rightarrow  \frac{\partial \vec{B}}{\partial t} = -\eta \frac{1}{\mu_{0}} \frac{\partial}{\partial \vec{r}} \left( \frac{\partial \vec{B}}{\partial \vec{r}}  \right) $$
If we follow the Gaussian profile decay in time we can measure the decrease in the magnetic field module by the numerical magnetic diffusivity as $\Delta B \approx \Theta \Delta t + ctte$, where

$$\Theta = -\eta \frac{1}{\mu_{0}} \frac{\partial}{\partial \vec{r}} \left( \frac{\partial \vec{B}}{\partial \vec{r}}  \right) $$
The next plot shows the decay of the Gaussian with the time for each model dimension.

\begin{figure}[h]
\centering
\includegraphics[width=0.3\textwidth]{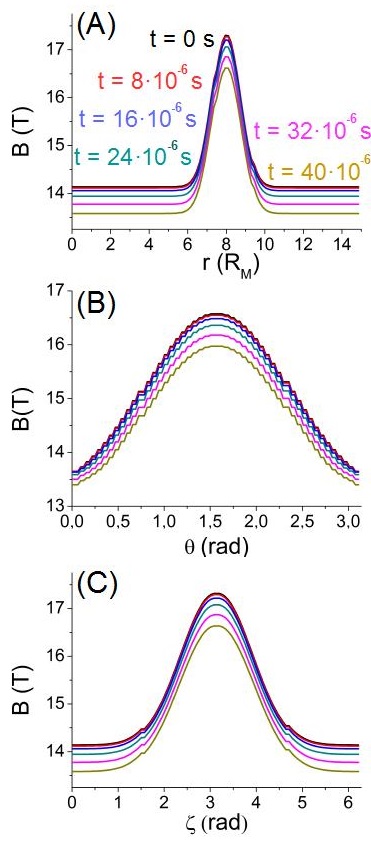}
\caption{Gaussian decay with time in the radial (a) and angular (b and c) directions at six different times.}
\label{12}
\end{figure} 

The value of the decay rate for each model dimension is obtained from the slope of the linear regression: $\Theta_{r} = 4.65 \cdot 10^{-5}$, $\Theta_{\theta} = 4.86 \cdot 10^{-5}$,  and $\Theta_{\zeta} = 4.29 \cdot 10^{-5}$. Next, the second derivative of the magnetic field is calculated at different time steps,  as is the numerical magnetic diffusivity. The numerical magnetic diffusivity is defined as the average of the values obtained: $\eta_{i} = (8.02 , 7.48 , 6.60) \cdot 10^{-3}$ m$^2$/s with $i=r$, $\theta$, $\zeta$. If we re-scale these values to that of the planet's scale (now assuming a characteristic length scale $L$ of $10^{6}$ m), we obtain for our setup $|\eta| \approx 1.81 \cdot 10^{8}$ m$^2$/s and $R_{m} = 1350$.
 
\section{Summary of simulation parameters:}

\begin{table*}[h]
\centering
\begin{tabular}{c | c c c c c c c}
Model & $B_{SW}$ (nT) & tilt ($^{o}$) & $|B_{dip}|$ (nT) & $|B_{quad}|$ (nT) & $R_{MP}/R_{ex}$ & $R_{magl}/R_{ex}$ & $R_{magw}/R_{ex}$ \\ \hline
B250 Bx & $(20, 0, 0)$ & $0$ & $250$ & $0$ & $0.33$ & $10.8$ & $2.3$ \\
B250 Bxneg & $(-20, 0, 0)$ & $0$ & $250$ & $0$ & $0.32$ & $10.8$ & $2.2$ \\
B250 By & $(0, 20, 0)$ & $0$ & $250$ & $0$ & $0.32$ & $9.2$ & $4.1$ \\
B250 Byneg & $(0, -20, 0)$ & $0$ & $250$ & $0$ & $0.32$ & $9.2$ & $4.3$ \\
B250 Bz & $(0, 0, 20)$ & $0$ & $250$ & $0$ & $0.50$ & $14.8$ & $5.1$ \\
B250 Bzneg & $(0, 0, -20)$ & $0$ & $250$ & $0$ & $0.00$ & $11.3$ & $3.2$ \\
B1000 Bx & $(20, 0, 0)$ & $0$ & $1000$ & $0$ & $1.11$ & $22.0$ & $2.2$ \\
B1000 Bxneg & $(-20, 0, 0)$ & $0$ & $1000$ & $0$ & $1.11$ & $22.8$ & $2.9$ \\
B1000 By & $(0, 20, 0)$ & $0$ & $1000$ & $0$ & $1.06$ & $17.5$ & $6.6$ \\
B1000 Byneg & $(0, -20, 0)$ & $0$ & $1000$ & $0$ & $1.06$ & $17.7$ & $6.5$ \\
B1000 Bz & $(0, 0, 20)$ & $0$ & $1000$ & $0$ & $1.29$ & $28.9$ & $9.9$ \\
B1000 Bzneg & $(0, 0, -20)$ & $0$ & $1000$ & $0$ & $0.87$ & $22.0$ & $1.9$ \\
B6000 Bx & $(20, 0, 0)$ & $0$ & $6000$ & $0$ & $3.65$ & $73.2$ & $22.9$ \\
B6000 Bxneg & $(-20, 0, 0)$ & $0$ & $6000$ & $0$ & $3.50$ & $73.4$ & $23.2$ \\
B6000 By & $(0, 20, 0)$ & $0$ & $6000$ & $0$ & $3.50$ & $56.1$ & $22.4$ \\
B6000 Byneg & $(0, -20, 0)$ & $0$ & $6000$ & $0$ & $3.55$ & $49.4$ & $19.8$ \\
B6000 Bz & $(0, 0, 20)$ & $0$ & $6000$ & $0$ & $4.11$ & $74.3$ & $26.4$ \\
B6000 Bzneg & $(0, 0, -20)$ & $0$ & $6000$ & $0$ & $3.31$ & $56.8$ & $21.6$ \\
Q02 Bx & $(20, 0, 0)$ & $0$ & $4800$ & $1200$ & $2.75$ & $40.6$ & $19.2$ \\
Q02 Bxneg & $(-20, 0, 0)$ & $0$ & $4800$ & $1200$ & $2.65$ & $41.5$ & $22.2$ \\
Q02 By & $(0, 20, 0)$ & $0$ & $4800$ & $1200$ & $2.65$ & $37.2$ & $17.5$ \\
Q02 Byneg & $(0, -20, 0)$ & $0$ & $4800$ & $1200$ & $2.64$ & $37.5$ & $16.6$ \\
Q02 Bz & $(0, 0, 20)$ & $0$ & $4800$ & $1200$ & $2.83$ & $34.2$ & $21.4$ \\
Q02 Bzneg & $(0, 0, -20)$ & $0$ & $4800$ & $1200$ & $2.58$ & $33.5$ & $17.8$ \\
Q04 Bx & $(20, 0, 0)$ & $0$ & $3600$ & $2400$ & $1.59$ & $49.5$ & $5.7$ \\
Q04 Bxneg & $(-20, 0, 0)$ & $0$ & $3600$ & $2400$ & $1.64$ & $49.8$ & $5.6$ \\
Q04 By & $(0, 20, 0)$ & $0$ & $3600$ & $2400$ & $1.62$ & $41.9$ & $7.1$ \\
Q04 Byneg & $(0, -20, 0)$ & $0$ & $3600$ & $2400$ & $1.66$ & $42.3$ & $6.9$ \\
Q04 Bz & $(0, 0, 20)$ & $0$ & $3600$ & $2400$ & $1.55$ & $49.8$ & $9.3$ \\
Q04 Bzneg & $(0, 0, -20)$ & $0$ & $3600$ & $2400$ & $1.68$ & $46.6$ & $4.6$ \\
tilt30 Bx & $(20, 0, 0)$ & $30$ & $6000$ & $0$ & $3.82$ & $144.5$ & $33.6$ \\
tilt30 Bxneg & $(-20, 0, 0)$ & $30$ & $6000$ & $0$ & $3.95$ & $101.5$ & $33.5$ \\
tilt30 By & $(0, 20, 0)$ & $30$ & $6000$ & $0$ & $4.18$ & $44.7$ & $13.2$ \\
tilt30 Byneg & $(0, -20, 0)$ & $30$ & $6000$ & $0$ & $4.41$ & $43.3$ & $13.1$ \\
tilt30 Bz & $(0, 0, 20)$ & $30$ & $6000$ & $0$ & $3.77$ & $39.5$ & $15.8$ \\
tilt30 Bzneg & $(0, 0, -20)$ & $30$ & $6000$ & $0$ & $3.94$ & $59.8$ & $19.2$ \\
tilt60 Bx & $(20, 0, 0)$ & $60$ & $6000$ & $0$ & $4.48$ & $81.6$ & $17.3$ \\
tilt60 Bxneg & $(-20, 0, 0)$ & $60$ & $6000$ & $0$ & $4.34$ & $100.1$ & $16.4$ \\
tilt60 By & $(0, 20, 0)$ & $60$ & $6000$ & $0$ & $4.49$ & $42.3$ & $6.6$ \\
tilt60 Byneg & $(0, -20, 0)$ & $60$ & $6000$ & $0$ & $4.43$ & $38.9$ & $5.6$ \\
tilt60 Bz & $(0, 0, 20)$ & $60$ & $6000$ & $0$ & $4.74$ & $47.3$ & $14.1$ \\
tilt60 Bzneg & $(0, 0, -20)$ & $60$ & $6000$ & $0$ & $4.53$ & $67.0$ & $12.9$ \\
tilt90 Bx & $(20, 0, 0)$ & $90$ & $6000$ & $0$ & $0.00$ & $30.3$ & $4.1$ \\
tilt90 Bxneg & $(-20, 0, 0)$ & $90$ & $6000$ & $0$ & $0.00$ & $54.9$ & $4.0$ \\
tilt90 By & $(0, 20, 0)$ & $90$ & $6000$ & $0$ & $0.00$ & $32.5$ & $2.5$ \\
tilt90 Byneg & $(0, -20, 0)$ & $90$ & $6000$ & $0$ & $0.00$ & $31.2$ & $2.6$ \\
tilt90 Bz & $(0, 0, 20)$ & $90$ & $6000$ & $0$ & $0.00$ & $55.1$ & $3.3$ \\
tilt90 Bzneg & $(0, 0, -20)$ & $90$ & $6000$ & $0$ & $0.00$ & $35.8$ & $4.4$ \\

\end{tabular}
\caption{Summary of simulation parameters.}
\label{table7}
\end{table*} 
 
The parameters $R_{magl}/R_{ex}$ and $R_{magw}/R_{ex}$ in Table B.1 are the magnetotail length and width normalized to the exoplanet radius.
 
\end{appendix}


\end{document}